\PassOptionsToPackage{left=2cm,right=1.5cm,top=1.5cm,bottom=1.5cm}{geometry}

\documentclass[pdflatex,sn-basic]{sn-jnl}

\usepackage{graphicx}%
\usepackage{multirow}%
\usepackage{amsmath,amssymb,amsfonts}%
\usepackage{amsthm}%
\usepackage{mathrsfs}%
\usepackage[title]{appendix}%
\usepackage{xcolor}%
\usepackage{textcomp}%
\usepackage{manyfoot}%
\usepackage{booktabs}%
\usepackage{algorithm}%
\usepackage{algorithmicx}%
\usepackage{algpseudocode}%
\usepackage{listings}%

\usepackage[export]{adjustbox}
\usepackage{longtable}
\usepackage{placeins}
\usepackage{url}
\usepackage{threeparttable} 

\theoremstyle{thmstyleone}%
\theoremstyle{thmstyletwo}%

\theoremstyle{thmstylethree}%

\begin{document}

\title[Topological and Temporal Stability Analysis of the Lightning Network]{Topological and Temporal Stability Analysis of the Lightning Network}


\author*[1,2]{\fnm{Danila} \sur{Valko}}\email{danila.valko@offis.de}

\author[1,2]{\fnm{Jorge} \sur{Marx G\'omez}}\email{jorge.marx.gomez@uol.de}


\affil*[1]{\orgdiv{Fakultät II - Department für Informatik}, \orgname{Carl von Ossietzky Universität Oldenburg}, \orgaddress{\street{Ammerländer Heerstraße, 114-118}, \city{Oldenburg}, \postcode{26129}, \state{Lower Saxony}, \country{Germany}}}

\affil[2]{\orgdiv{Energy Division}, \orgname{OFFIS – Institute for Information Technology}, \orgaddress{\street{Escherweg, 2}, \city{Oldenburg}, \postcode{26121}, \state{Lower Saxony}, \country{Germany}}}



\abstract{The Lightning Network (LN) is the most prominent payment channel network built atop Bitcoin, designed to enable scalable, low-cost off-chain transactions. Understanding its structural evolution and temporal stability is critical for routing optimization, liquidity allocation, and infrastructure robustness. Leveraging a validated dataset of LN topology snapshots spanning 2019–2023, we compute a set of network-science metrics under directed, undirected, unweighted, capacity-weighted, and routing-aware graph representations. To our knowledge, this is the first longitudinal multi-representation temporal stability analysis of the Lightning Network combining topological, distributional, and routing-equivalent persistence metrics over a 5-year validated snapshot dataset.

Developed analytical framework reveals a network undergoing gradual structural sparsification and increasing modularization: density, clustering, and global efficiency decline over time, while community fragmentation and centralization persist. However, distributional and observed operational characteristics remain remarkably stable. Degree and shared node capacity distributions exhibit relatively small temporal deviation according to Kolmogorov–Smirnov and Wasserstein diagnostics. The average node retention remains above 90\% between successive snapshots. While the average channel retention exceeds 90\% in the unweighted representation, it decreases to approximately 70\% when evaluated using more realistic pathfinding strategies (LND, CLN, and ECL). Despite this reduction, observed routing-equivalent payment paths remain largely preserved over time.

These results indicate that the LN is structurally reconfiguring yet operationally persistent: mesoscopic cohesion decreases, but the routing backbone and economic influence structure remain stable. At the same time, extreme betweenness centralization suggests a growing concentration of routing influence that may increase vulnerability to correlated failures, although explicit robustness under adversarial conditions was not evaluated in this study.

Beyond empirical findings, this work provides a comprehensive characterization of a reproducible benchmark dataset and a temporal–topological stability analysis framework for payment channel networks, supporting future research in robustness analysis, routing optimization, and decentralized infrastructure design.}

\keywords{Lightning Network, network topology, payment channel networks, temporal network analysis, topological stability}



\maketitle

\maketitle

\section{Introduction}
\label{sec:introduction}
Payment Channel Networks (PCNs) are off-chain scaling solutions that enable users to transact through pre-funded, bidirectional channels without broadcasting each transaction to the underlying blockchain. By settling only the final results on-chain, PCNs significantly reduce network congestion and transaction fees\citep{DIVAKARUNI2023}, thus enabling high-throughput and low-cost micro payments. The Lightning Network (LN) is the most prominent implementation of a PCN, built atop Bitcoin. 
LN holds significant promise as a scalable and efficient layer for decentralized digital payments\citep{Lin2022, DIVAKARUNI2023}. However, realizing its full potential requires a deeper understanding of its network topology and systemic behaviour. Ongoing development efforts are focused on improving routing efficiency, enhancing privacy, and expanding liquidity through techniques such as channel splicing and multi-path payments\citep{ValkoMarxGomezReview2025}. Despite this progress, many challenges persist – particularly in the areas of infrastructure utilization, channel management\citep{Herrera, Rohrer2019}, and pathfinding efficiency\citep{ValkoMarxGomezReview2025, ValkoKudenko2024}.

\subsection{Background and related work}
The topology of the LN was first systematically analysed by Seres et al.\citep{Seres2020}. Their research, studied two network snapshots from January 2019, introduced early insights into degree distribution and explored the network's robustness against random node failures. A more detailed investigation was conducted by Rohrer et al.\citep{Rohrer2019}, who also assessed robustness and introduced key topological metrics such as network diameter and average distance. Their analysis, based on two observation periods (October-November 2018 and January-February 2019), emphasized structural vulnerabilities associated with node dominance. Subsequently, Martinazzi and Flori\citep{Martinazzi2020} presented an early study covering one year of the LN's evolution from January 2, 2018, to January 12, 2019. They focused on centralization trends, some efficiency metrics, and the network's resilience against attacks exploiting the influence of highly connected nodes.

Expanding the analytical scope, Zabka et al.\citep{Zabka2021, zabka2022} conducted a large-scale empirical study, classifying LN nodes by software client and geographic location using seven snapshots between March 2018 and January 2020. Their analysis revealed that node proximity played a minor role, so that nodes tend to connect to large hubs, even if there is a large geographical distance between them. Their accurate client classification contributed valuable insights into the LN's implementation diversity, which is relevant for economical analysis and assessing security risks. Together with Mizrahi and Zohar\citep{Mizrahi2021}, these works laid the foundation for accessing basic LN topology characteristics, although their analyses were limited in temporal scope.

The most comprehensive and up-to-date study is also done by Zabka et al. (2024)\citep{Zabka2024}, who analysed ten reconstructed network snapshots from 2019 to 2021. Their work revealed that the top 5\% of nodes handle the majority of transaction paths, and that centrality inequality (measured by the \textit{Gini index}) increased by more than 15 points during the observed period. These findings suggest growing structural centralization, raising critical concerns about the network's resilience, fairness, and user privacy. Complementing this empirical approach, Guasoni et al.\citep{Guasoni2024} explored LN topologies from a theoretical standpoint. They demonstrated that minimizing routing costs can inherently lead to centralized structures. Their work showed that finding the globally optimal topology is NP-complete and that intuitive configurations, such as star topologies, are not always cost-optimal.

Despite covering broader time windows and introducing new topology-related metrics, current research still provides only a partial view of what full-scale network science analysis could offer. A more holistic and temporally consistent approach is needed to fully capture the dynamic structural properties of the LN.

From a research standpoint, access to validated and well-documented network data is critical for accurate analysing, modelling, benchmarking, and simulation of real-world LN behaviour. Notably, there exists a single major public data source that captures LN gossip messages over a long time period, now 2019-2023\citep{lngossip}. These messages can be used to reconstruct historical snapshots of the LN topology at different points in time. While this data has supported research in areas such as routing optimization and privacy and liquidity management, only a limited number of studies have deeply examined the architectural, topological, and temporal features of the LN itself (e.g., \citep{Seres2020, Zabka2021, zabka2022, Zabka2024}). To date, there has been no comprehensive analysis utilizing this data source, nor any work thoroughly characterizing the network's structural evolution and stability trends using a wide set of network science metrics.

This gap is significant, particularly for applications relying on heuristic-based pathfinding, routing optimization, or pattern mining – each of which assumes some degree of structural consistency and topological stability in the underlying network\citep{ValkoKudenko2024, ValkoKudenko2025}. Discovering the temporal stability of LN's topology is thus essential. 

To this end, we previously developed and validated a dataset containing 336 Lightning Network topology snapshots\citep{dataset2025, datapaper2025} spanning the years 2019–2023, which we use here to conduct a comprehensive network analysis. 

This work equips developers with validated, computationally intensive network-science metrics and topology benchmarks to support rigorous testing and algorithm design, while advancing PCN research and promoting reproducible LN studies. Moreover, it introduces a novel stability analysis framework grounded in temporal–topological backbone measures, enabling systematic assessment of node and channel retention, routing capacity preservation, and long-term observed operational feasibility.

\subsection{Contribution and practical implications}

This paper makes a twofold contribution. First, we propose a novel stability analysis framework for LN-like PCNs, based on temporal-topological stability measures of the network backbone. The framework evaluates: (i) the technical retention of nodes and channels over time; (ii) routing stability, considering the preservation of channel capacity over time; and (iii) observable operational stability, considering cost-based over time payment feasibility across the network.

Second, we contribute to both the network science and LN developer communities through a comprehensive characterization of a developed and validated dataset comprising 336 LN topology snapshots. A diverse set of computationally intensive network-science metrics is shared in machine-readable format. It serves multiple purposes: (i) providing analytically sound reference characteristics of the LN over time, which are essential for future research on LN topology, dynamics, and evolution, complemented by an interactive data dashboard (\url{https://ellariel.github.io/ln-data-dashboard/}); (ii) enabling comprehensive topology analysis with sensitivity checks and uncertainty quantification, under both directed and undirected graph representations, as well as realistic routing strategies based on the native LN pathfinding algorithms (\textit{LND}\citep{LND}, \textit{CLN}\citep{CLN}, \textit{ECL}\citep{ECL}); and (iii) supporting informed network snapshot selection and sampling for developers working on network robustness algorithms and heuristics who require realistic (rather than synthetic) benchmarks and validated assumptions -- this addresses a well-known practical gap in the field\citep{ValkoMarxGomezReview2025}.

Since undirected and unweighted graph representations are a common simplification in LN topology analysis (see, e.g., \citep{Wang2022, Lin2022}), primarily because they are less computationally expensive and because many topology focused metrics do not require weighted or directed structures, our extensive sensitivity analysis implicitly also addresses whether this simplified approach is analytically appropriate.

\section{Data and methods}
\label{sec_methods}

\subsection{Data}

\subsubsection{Data source and preparation pipeline}
\label{sec_methods_sub1}

The dataset comprises 336 snapshots of the LN topology reconstructed from publicly available gossip message archives collected between 2019 and 2023. It was created through a multi-stage pipeline, for details, see \citep{datapaper2025}. 

The dataset construction followed four main stages, transforming raw gossip messages into a structured, geolocated, and analysis-ready network graphs. First, in the data acquisition stage, over 35 million LN gossip messages, originally broadcasted according to the Basics of Lightning Technology (BOLT) protocol and archived by Christian Decker and collaborators, were collected from seven historical archives\citep{lngossip}. These raw messages describe node metadata, channel creation, and directional channel parameters, forming the demi-structured raw input for the pipeline.

Second, in the network graph reconstruction stage, the \textit{TimeMachine} algorithm\citep{Zabka2024} was used to replay, deduplicate, and chronologically order gossip messages up to specific timestamps. Through this controlled replay process, raw announcements were transformed into coherent historical topology states: nodes became graph vertices with associated metadata, and channels became bidirectional edges with direction-specific attributes. Reconstructing these states at regular two-week intervals produced a large set of temporal network slices approximating the visible LN topology at each available date. The available metadata for nodes and channels is described in our data paper\citep{datapaper2025}.

Third, during pre-processing and consistency control, overlapping or inconsistent reconstructions were resolved by retaining the most connected graph per interval. Isolated nodes and channels lacking sufficient metadata were removed, ensuring structural completeness and analytical validity. This step converted the reconstructed but potentially noisy graphs into 336 internally consistent topology snapshots spanning January 2019 to July 2023.

Finally, the dataset was enriched with geographical metadata. The resulting final dataset, including \textit{NetworkX}-compatible GML files and a summary metadata in the CSV format, was publicly hosted on the Harvard Dataverse (available at \url{https://doi.org/10.7910/DVN/2OAVO6}).

\subsubsection{Data coverage and validity}
\label{sec_methods_sub2}

\begin{figure*}[t]
\hspace{-2.2cm}
\includegraphics[scale=0.55, center]{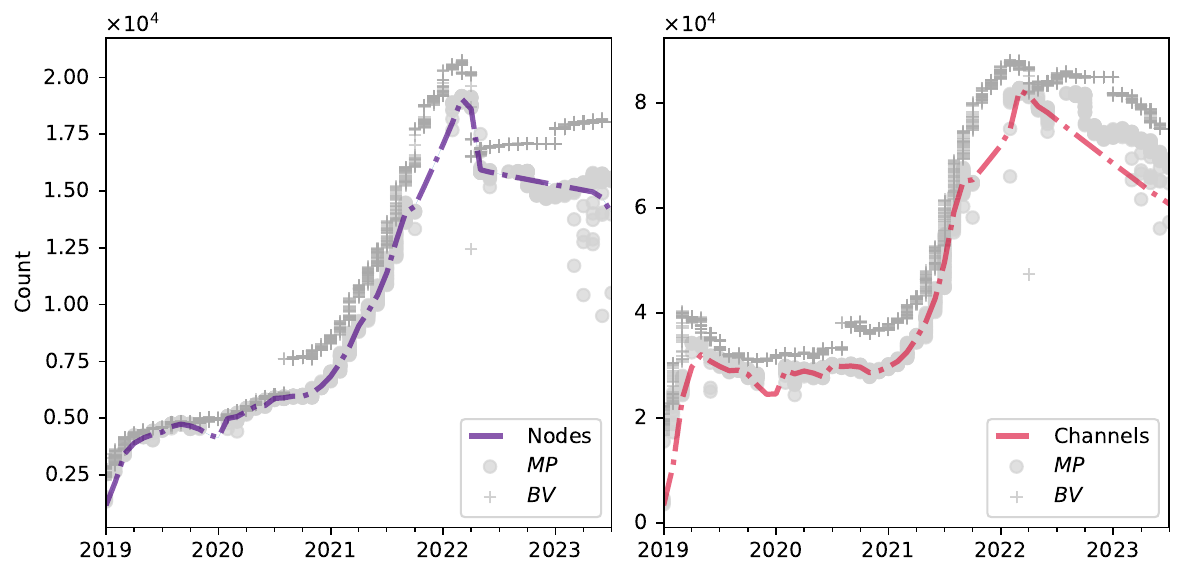}
\hspace{-2.2cm}
\includegraphics[scale=0.4, center]{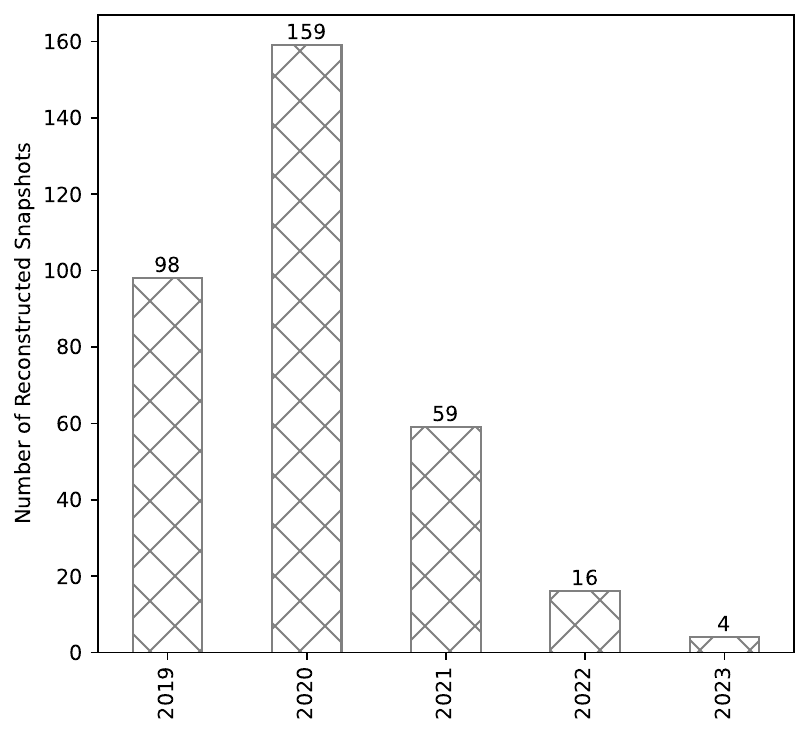}
\caption{Node and channel quantity and data coverage over time\citep{datapaper2025}.}
\label{fig01}

\begin{tablenotes}[]\footnotesize\item[]
Note. Snapshot coverage (bottom figure) and monthly averaged size measurements (top figures) from reference sources (\textit{BV}, \textit{MP} with dots and crosses, respectively) and the reconstructed data snapshots (interpolated dashed lines), \textit{BV} – BitcoinVisuals (\url{https://bitcoinvisuals.com/lightning}), \textit{MP} – mempool.space (\url{https://mempool.space/graphs/lightning/nodes-networks}).
\end{tablenotes}

\end{figure*}

The dataset spans January 2019 to July 2023, capturing the temporal evolution of the LN over nearly five years with irregular but dense snapshot coverage, see, \textit{Figure~\ref{fig01}}. Because gossip messages are not officially archived by the LN protocol, temporal gaps arise from uneven message availability, asynchronous propagation, and periodic changes in collection methodology; these constraints are quantitatively documented in the published dataset metadata\citep{dataset2025} and subsequent data paper\citep{datapaper2025}, therefore they were not described here in full.

As described in the data paper\citep{datapaper2025}, to assess validity, cross-comparisons of network size characteristics (nodes, channels, graph diameter) were conducted against independent reference sources including \textit{BitcoinVisuals} and \textit{mempool.space} statistics. Resulting statistical tests (Kolmogorov–Smirnov test, Wasserstein distance, Spearman's correlation) indicated strong consistency between the reconstructed dataset and validation measurements, supporting the validity despite data collection limitations. Spatial coverage quality was further evaluated by comparing proportions of geolocated nodes against independent estimates of publicly visible LN nodes, showing alignment with known trends (root mean square deviation of 0.57\% and the correlation coefficient with reference values is 0.98). Remaining limitations, including non-uniform temporal coverage are transparently reported alongside the dataset, see, \citep{datapaper2025}.

\subsection{Analytical strategy and tools}
\subsubsection{Network backbone topology analysis}

For the purpose of this work we selected several key metrics used in conventional network analysis\citep{Newman2003} that address network structure, connectivity, resilience, emergent patterns etc., see, \textit{Table~\ref{tab:ln_selected_metrics}}. However, in order to provide 
comprehensive reference characteristics of the curated benchmark dataset and for the sake of enabling reproducible future research on LN topology, dynamics, and evolution, in total 47 attributes and metrics were computed with some analytical redundancy, see full set of measurements in \textit{Table~\ref{tab:ln_metrics_summary}}. Therefore, this diverse set of computationally intensive network-science metrics is shared in machine-
readable format and complemented by an interactive
data dashboard (available at \url{https://ellariel.github.io/ln-data-dashboard/}).

We used the Python library \textit{NetworkX}\citep{SciPyProceedings} to calculate most metrics, while the rest were coded manually. All scripts are available in an open repository (available at \url{https://github.com/ellariel/ln-comprehensive-analysis}).

\newpage

\begin{table*}[!htbp]
\caption{Full set of metrics and attributes for network analysis.}
\label{tab:ln_metrics_summary}
\centering
\small
\begin{tabular}{|p{.21\textwidth}|p{.45\textwidth}|p{.21\textwidth}|}
\hline
\textbf{Category} & \textbf{Metrics and attributes} & \textbf{Focus} \\
\hline
\textbf{Network Structure}: Basic Topology Measures, Assortativity \& Degree & 
\textit{nodes}, \textit{edges (channels)}, \textit{components}, \textit{density}, \textit{diameter}, \textit{shortest path length}, \textit{average degree}, \textit{degree assortativity}\citep{Newman}
& Describe overall architecture, scale, and connectivity patterns of the network. \\
\hline
\textbf{Connectivity \& Resilience}: Connectivity, Clustering \& Transitivity &
\textit{bridges}, \textit{average node connectivity}\citep{BEINEKE2002}, \textit{minimal edge cover}, \textit{transitivity}\citep{PERON20231}, \textit{average clustering}\citep{SchankWagner2004}
& Measure cohesion, redundancy, and vulnerability to failure. \\
\hline
\textbf{Function \& Dynamics}: Efficiency \& Info Flow, Centrality, Structural Holes &
\textit{global efficiency}\citep{Latora}, \textit{information centrality}\citep{Brandes2005CentralityMB}, \textit{average betweenness centrality}\citep{BRANDES2008136}, \textit{communicability betweenness centrality}\citep{ESTRADA2009764}, \textit{common neighbour centrality}\citep{Ahmad2020}, \textit{constraint value}, \textit{effective size}\citep{BURTRonaldS2004SHaG, Lutter2017}, \textit{Burt's effective size}\citep{Lutter2017}, \textit{closeness vitality}
& Analyze information/value flow and identify nodes with control or influence roles. \\
\hline
\textbf{Emergent Patterns}: Link Prediction \& Community Detection &
\textit{resource allocation index}\citep{Zhou}, \textit{Jaccard coefficient}\citep{Nowell}, \textit{preferential attachment}\citep{Nowell}, communities (\textit{FLP}\citep{Traag2023}, \textit{ALP}\citep{PhysRevE}, \textit{GM}\citep{ClausetNewmanMoore2004})
& Detect underlying community structure and predict potential future links. \\
\hline
\textbf{Other Metrics}: Topological Stability, Centrality Inequality etc. &
 \textit{Gini betweenness centrality}\citep{Zabka2024}, \textit{Wiener index}\citep{Mircea}, \textit{Wasserstein distance}\citep{Ramdas}, \textit{degree and sared capcity distribution approximations}, \textit{degree distribution entropy}\citep{Jones2022}, \textit{node retention rates}, \textit{channel retention rates}, \textit{Kolmogorov–Smirnov statistics}
& Analyze specific payment channel network features. \\
\hline
\end{tabular}
\end{table*}

\begin{table*}[!htbp]
\caption{Key metrics suitable for sensitivity check.}
\label{tab:ln_selected_metrics}
\centering
\small
\begin{tabular}{|p{.25\textwidth}|p{.15\textwidth}|p{.15\textwidth}|p{.15\textwidth}|p{.15\textwidth}|}
\hline
\textbf{Key metrics} & \textbf{Undirected} & \textbf{Directed} & \textbf{Unweighted} & \textbf{Weighted} 
\\\hline
\textit{density} & \checkmark & \checkmark & \checkmark &
\\\hline
\textit{bridges} & \checkmark & & \checkmark &
\\\hline
\textit{transitivity} & \checkmark & \checkmark & \checkmark &
\\\hline
\textit{average degree} & \checkmark & \checkmark & \checkmark &
\\\hline
\textit{average clustering} & \checkmark & \checkmark & \checkmark & \checkmark 
\\\hline
\textit{minimal edge cover} & \checkmark & & \checkmark &
\\\hline
\textit{global efficiency} & \checkmark & & \checkmark &
\\\hline
\textit{effective size} & \checkmark & & \checkmark & 
\\\hline
\textit{Burt's effective size} & \checkmark & & \checkmark & 
\\\hline
\textit{information centrality} & \checkmark & & \checkmark & \checkmark
\\\hline
\textit{Jaccard coefficient} & \checkmark & & \checkmark &
\\\hline
\textit{preferential attachment score} & \checkmark & & \checkmark &
\\\hline
\textit{resource allocation index} & \checkmark & & \checkmark &
\\\hline
\textit{communities (\textit{FLP}, \textit{GM})} & \checkmark & \checkmark & \checkmark & \checkmark
\\\hline
\textit{Gini betweenness centrality} & \checkmark & \checkmark & \checkmark & \checkmark
\\\hline
\textit{degree distribution} and \textit{shared capacity distribution statistics} & \checkmark & \checkmark & & (\checkmark)
\\\hline
\textit{node retention rate} and \textit{shared node capacity rate} & \checkmark &  \checkmark & \checkmark & (\checkmark)
\\\hline
\textit{channel retention rate} & \checkmark & \checkmark & \checkmark & \checkmark
\\\hline
\end{tabular}
\end{table*}

\subsubsection{Sensitivity and uncertainty considerations}

Despite undirected graph representation is a common simplification in the LN analysis (see, e.g., \citep{Wang2022, Lin2022}) several methods appear to be sensitive to directed graph representation as it may affect average degree and pathfinding strategies. Thus we decided to test key metrics of reported topology analysis that are defined and implemented also for undirected graphs, see them in \textit{Table~\ref{tab:ln_selected_metrics}}.

Some metrics are also sensitive to weighted graph representation as it affects connectivity and flow heuristics\citep{Lin2022}, as well as the efficiency of pathfinding strategies. In LN this relates to three cases: (i) unweighted graph representation, which reveals technical or backbone topology characteristics calculated via shortest-path pathfinding algorithm; (ii) weighted by channel capacity representation, which reflects operational reality of the network rather than just technical channel feasibility; (iii) weighted by pathfinding cost representation, which relies on technical-economical channel characteristics (such as fees, locktimes, capacities etc.) and cost-based channel feasibility across the network, which is important in practice. 

Thus, to better reflect realistic pathfinding strategies and the operational environment of the LN, we considered two generic weighting schemes in addition to the simple unweighted case. In the baseline or default scenario, (i) all weights are equal:
\begin{equation} \label{eq1}
\begin{split}
w_{DEF} = 1.
\end{split}
\end{equation}

We then applied (ii) a capacity-based weighting scheme for the network represented by a graph $G=(V,E)$, where the weight of an existing edge (representing a payment channel) $[u,v] \in E$ between vertices (representing payment nodes) $u, v \in V$ reflects channel's normalized capacity ($capacity_{norm}$) and defined as:
\begin{equation} \label{eq2}
\begin{split}
w_{CAP}[u,v] = 1 - capacity_{norm}[u,v].
\end{split}
\end{equation}

It is important to note that capacity serves as the only  appropriate proxy for a channel's economic value, since exact channel balances and payment transactions remain private\citep{Herrera}.

Finally, in the third scenario, (iii) weights approximate the exact cost functions used by native payment pathfinding algorithms.
Currently, three major LN software clients implement their own routing algorithms: LND\citep{LND}, c-Lightning (CLN)\citep{CLN}, and Eclair (ECL)\citep{ECL}. Although these implementations differ in design, each uses publicly available channel parameters to compute the cost of adding a channel $[u,v]$ to a payment path. In our analysis, we relied on their existing analytical implementations\citep{R117} to derive the corresponding weights of interest.

In particular, the LND cost function (used as the weight function or just weight, $w_{LND}$) accounts for the payment amount, the time-lock delay ($locktime$), and routing fees. Additionally, it incorporates a penalty term that biases the algorithm against channels with recently observed failures.

\begin{equation} \label{eq3}
\begin{split}
w_{LND}[u,v]&=amount[u,v]\cdot locktime[u,v]\cdot\rho \\
&+ fee[u,v]+bias[u,v],
\end{split}
\end{equation}
here $amount$ is a payment size, $locktime$ – channel locktime, $fee$ – total fee paid, $\rho$ – a risk factor set to $15\cdot10^{-9}$ by default\citep{R117} and $bias$ accounts for the previous payment failures caused by the channel $[u,v]$. The value of $bias$ is extremely large during the first hour after failure and decreases exponentially with every hour elapsed after the last failure. For a more detailed description of the LND cost function see \citep{AndreescuRoosErsoy2021}.

In contrast, the CLN algorithm introduces some randomness in the path selection, which decreases both predictability and the probability of selecting a route that leads to failure repeatedly\citep{AndreescuRoosErsoy2021}. The randomness is introduced via $fuzz$ parameter, which is set to $0.05$ by default, and then a scaling factor is calculated: $scale = 1 + random(-fuzz, fuzz)$\citep{R117}. So that, 
\begin{equation} \label{eq4}
\begin{split}
w_{CLN}[u,v]&=(amount[u,v]+ scale\cdot fee[u, v]) \\
&\cdot locktime[u,v]\cdot\rho+bias[u,v].
\end{split}
\end{equation}

In addition to the fee, the ECL considers the normalized locktime ($locktime_{norm}$), capacity ($capacity_{norm}$) and the $age[u, v]$ of a channel in blocks as parameters that impact its cost\citep{R117}. So that, 
\begin{equation} \label{eq5}
\begin{split}
w_{ECL}[u,v]&=fee[u,v]\cdot (locktime_{norm}[u,v]\cdot locktime_{ratio} \\
&+ (1-capacity_{norm}[u,v])\cdot capacity_{ratio} \\
&+ n_{age}[u,v]\cdot age_{ratio}).
\end{split}
\end{equation}

Randomness is introduced by not necessarily choosing the path of the lowest cost. Rather, it chooses among the $k$-cheapest paths based on Yen's $k$ shortest paths algorithm. The default value of $k$ is set to be $3$\citep{R117}. ECL further utilizes default weights $locktime_{ratio} = 0.15$, $capacity_{ratio} = 0.5$, and $age_{ratio} = 0.35$ for the impact of locktime, capacity, and age, respectively\citep{R117}.

To ensure a comprehensive analysis, these sensitivities in graph structure and path-cost weighted representations must be incorporated into the analytical workflow. Accordingly, we identified the key metrics to be computed with additional sensitivity checks (see, \textit{Table~\ref{tab:ln_selected_metrics}}). To investigate topological and temporal trends in a transparent and interpretable manner, we report and analyse results for the available directed and weighted graph representations together, rather than separating them into an additional section.

Although our analysis relies on substantially broader temporal coverage than previous studies, the available snapshots are not uniformly distributed across time. In particular, observations are denser during 2019–2020 than during 2022–2023 owing to historical gossip archive availability. Consequently, estimates corresponding to sparsely sampled periods are associated with greater uncertainty. To mitigate this effect, all temporal summaries are reported as quarterly averages with bootstrapped 95\% confidence intervals, while the figures explicitly encode snapshot coverage through line intensity. Importantly, the principal temporal trends (e.g., decreasing density, clustering, transitivity, and global efficiency) remain consistent after quarterly aggregation, indicating that the reported patterns are not driven only by unequal sampling density, although the magnitude of late-period estimates should be interpreted with appropriate caution. For the sake of reproducibility, all random seeds were fixed; their values are provided in the accompanying repository.

\subsubsection{Structural and operational stability analysis}

To discover the temporal–topological stability of the LN, we propose a set of novel metrics. These metrics are designed to capture both the structural persistence of the network backbone over time and the transactional stability of its payment channels, which is crucial for the efficiency of payment pathfinding strategies. Specifically, the analysis evaluates: (i) the technical retention of nodes and channels across successive snapshots; (ii) temporal routing stability, through the preservation of the observed node and channel capacity over time; and (iii) operational stability, reflecting the cost-based observed feasibility of nodes within the evolving network.

To assess structural retention, we first define the \textit{node retention rate}, which utilises the fraction of overlapping nodes (based on unique node \textit{ID}s) between two successive snapshots at time steps \( t \) and \( t+1 \). Formally, for two graph snapshots \( G_t \) and \( G_{t+1} \), the node retention rate \( I_{node} \) is given by:
\begin{equation} \label{eq6}
\begin{split}
I_{node}({G_t, G_{t+1}}) = \frac{|V_t \cap V_{t+1}|}{|V_t|},
\end{split}
\end{equation}
where \( V_t \) and \( V_{t+1} \) denote the sets of nodes in snapshots \( G_t \) and \( G_{t+1} \), respectively.

Analogously, to quantify channel-level retention, we define the \textit{channel retention rate} \( I_{channel} \) as:

\begin{equation} \label{eq7}
\begin{split}
I_{channel}({G_t, G_{t+1}}) = \frac{|\tilde{E}_t \cap \tilde{E}_{t+1}|}{|\tilde{E}_t|},
\end{split}
\end{equation}
where \( \tilde{E}_t \) and \( \tilde{E}_{t+1} \) denote the sets of effective edges representing channels or equivalent payment paths defined on the node intersection \( V_t \cap V_{t+1} \), and that technically exist in snapshots \( G_t \) and \( G_{t+1} \). This formulation accounts for whether direct channels (or equivalent paths of equal or shorter length) exist between the same pair of nodes in both snapshots. In this way, the metric reflects potential network reconfigurations while preserving technical routing capability.

However, structural (technical) retention alone does not guarantee that the network remains economically operational over time. Therefore, we additionally examine the preservation of channel capacity and shared node capacity, which serve as proxies for the network’s economic stability.

We introduce the \textit{shared node capacity} for a graph snapshot $G=(V,E)$, defined as the sum of half-capacities of all channels $E_v \subseteq E$ incident to a node $v \in V$:
\begin{equation} \label{eq8}
\begin{split}
capacity_{node}^{shared}(v) = \frac{1}{2}\sum_{e \in E_v} capacity[e].
\end{split}
\end{equation}

The total shared node capacity of an arbitrary subset of nodes $S \subseteq V$, evaluated within a particular network snapshot $G$ is defined as $total~capacity_{node}^{shared}{}(S; G) = C_{total}(S; G) = \sum_{v \in S} capacity_{node}^{shared}(v)$. When calculating the total capacity of the entire graph, having $S=V$, this notation simplifies to $C_{total}(G)$. 

Note that, when comparing two successive snapshots at $t$ and $t+1$, the common node set is $S=V_t \cap V_{t+1}$. Importantly, the capacities are evaluated independently in each snapshot. Thus, $total~capacity_{node}^{shared}(S; G_t)$ or $C_{total}(S; G_t)$ denotes the total shared capacity of the common nodes using the channel capacities observed in snapshot $G_t$, whereas $total~capacity_{node}^{shared}(S; G_{t+1})$ or $C_{total}(S; G_{t+1})$ denotes the capacity of the same node set evaluated using the updated channel capacities in snapshot $G_{t+1}$.

Based on this, we define the \textit{shared node capacity rate} between two successive snapshots as:
\begin{equation} \label{eq9}
\begin{split}
I_{node}^{capacity}({G_t, G_{t+1}}) = \frac{C_{total}(V_t \cap V_{t+1}; G_t)}{C_{total}(G_t)} 
\cdot \frac{C_{total}(G_{t+1})}{C_{total}(V_t \cap V_{t+1}; G_{t+1})}.
\end{split}
\end{equation}

This metric evaluates whether the technically overlapping nodes preserve the same relative share of total network capacity across successive snapshots. In other words, it measures whether these nodes maintain a consistent impact on overall network capacity over time. The measure equals one if the total shared capacity remains proportionally unchanged between \( G_t \) and \( G_{t+1} \). It may exceed one if some shared channels are replaced by higher-capacity channels in the successor snapshot.

Finally, we define a weighted complement of the channel retention rate:
\begin{equation} \label{eq10}
\begin{split}
I_{channel}^{w}({G_t, G_{t+1}}) = \frac{|\tilde{E}_t^{w} \cap \tilde{E}_{t+1}^{w}|}{|\tilde{E}_t^{w}|},
\end{split}
\end{equation}
where \( \tilde{E}^{w} \) denotes the set of effective edges representing channels or equivalent payment paths defined on the node intersection \( V_t \cap V_{t+1} \), computed using the weighted graph representation and a pathfinding strategy with cost function represented by weight $w$. 

By definition, when $w = w_{DEF} = 1$, this measure reduces to its unweighted counterpart \eqref{eq7}: $I_{channel}({G_t, G_{t+1}}) = I_{channel}^{w_{DEF}}({G_t, G_{t+1}})$.

When capacity-respecting weights ($w_{CAP}$) are applied, the metric approximates so called a \textit{shared channel capacity rate} measure (as defined for nodes \eqref{eq9}), whereas alternative weighting schemes yield stability measures corresponding to the specific pathfinding strategies under consideration. Importantly, $I_{channel}^{w}$ measures preservation of observable routing capability considering equivalent paths of equal or lower total cost rather than exact channel identity. Thus, even when channels churn, functional routing equivalence may remain stable.

\section{Results}
\label{sec_results}

\subsection{Topology analysis}
\label{sec_results_sub1}

\subsubsection{Network sparsity}
The LN is widely recognized as a \textit{sparse} network, meaning that only a small fraction of all possible connections between nodes are actually present. This sparsity arises from its design as a \textit{peer-to-peer payment channel network}, where users open channels selectively, typically driven by considerations of performance, cost, or trust. Unlike fully connected or \textit{dense} networks, most LN nodes maintain relatively few connections, while a small subset of \textit{hub nodes} manage a disproportionately large number of channels\citep{Zabka2024}.

\begin{figure*}[t]
\centering
\includegraphics[scale=0.55, center]{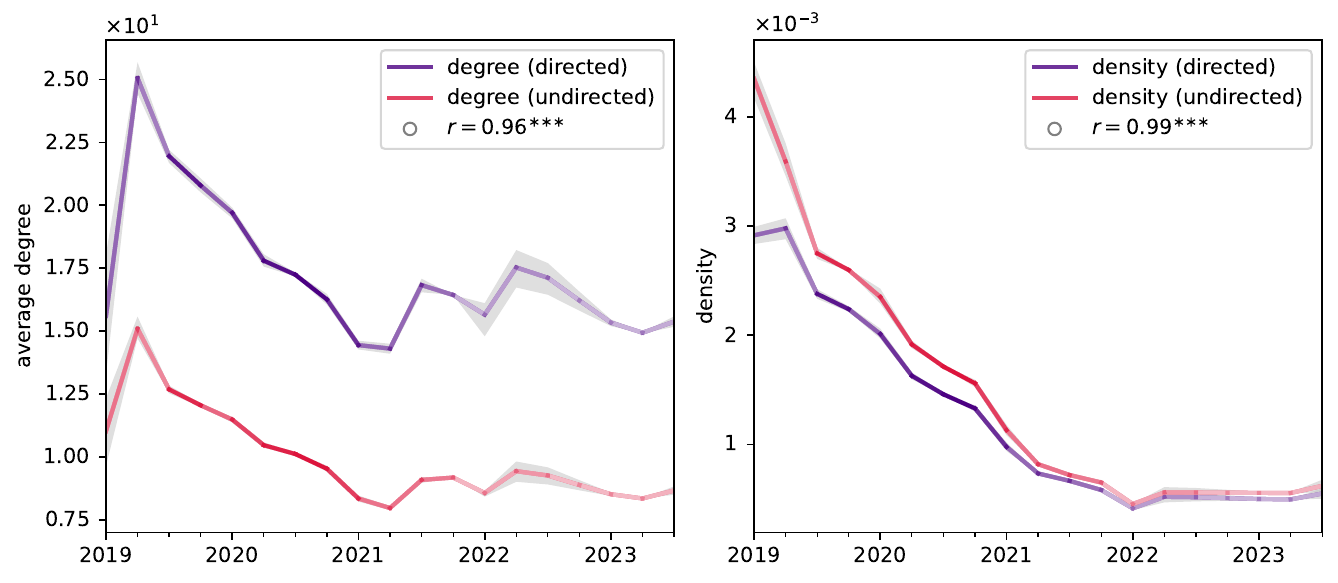}
\caption{The average node degree and the network density over time.}\label{fig02}
\begin{tablenotes}[]\footnotesize\item[]
Note. Quarterly averaged estimates (solid lines) are presented together with bootstrapped ($n = 100$) 95\% confidence intervals (shaded areas). Line colour intensity reflects the normalized data coverage for each time period based on the available snapshots. Significance levels for Spearman's correlation $p$-value: *** $< 0.001$, ** $< 0.01$, * $< 0.05$.
\end{tablenotes}
\end{figure*}

The principal indicator of sparsity is the low average node degree, where the majority of nodes have only a few connections, which is directly related to low edge density. For $n$ nodes, a fully connected graph contains $n(n-1) / 2$ edges. In contrast, the LN is expected to exhibit a relatively much lower channel count. Early studies already identified this low density and the resulting centralization around some hub nodes\citep{Rohrer2019, Seres2020}, an extended view on the current dynamics is on the \textit{Figure~\ref{fig02}}. 

Figure~\ref{fig02} illustrates that LN growth is accompanied by decreasing average connectivity per node. While node and channel counts increase, both average node degree and edge density decline, indicating that network expansion occurs primarily through the addition of sparsely connected nodes rather than densification of existing sub-graphs.

This sparsity has several important implications. First, routing efficiency relies heavily on a small number of central nodes, which can make the network more vulnerable to targeted attacks\citep{Mizrahi2021}. Second, the network benefits in terms of scalability and cost-efficiency, as fewer connections reduce overhead and allow for more flexible, on-demand channel creation. Finally, sparsity can improve privacy by limiting the visibility of transactions across the network, although actual privacy levels also depend on the specific routing algorithms used\citep{Papapetrou}. Importantly, declining density does not necessarily imply deteriorating reachability, as scale-free networks can maintain short path lengths despite sparsification. 

Having established that the LN is structurally sparse, we next examine whether this sparsity follows a particular pattern. Specifically, whether the LN's degree distribution exhibits the heavy-tailed behaviour characteristic of scale-free networks.

\subsubsection{Scale-free network properties}
A scale-free network is characterized by a degree distribution that follows a power law, at least asymptotically. Mathematically, the probability $P(k)$ that a node has degree $k$ follows $P(k) \sim k^{-\alpha}$, where $\alpha$ is typically between 2 and 3\citep{Clauset}. This distribution implies that while most nodes have few connections and a small number of nodes, that are known as \textit{hubs}, are highly connected and play a critical role in maintaining the structure and function of the network. Recent studies have shown both its persistent sparsity and the dominance of a small number of high-degree nodes in its topology\citep{Zabka2021, Zabka2024}, but we want to confirm this over a broad time frame and also highlight the trend.

The key characteristics of a scale-free network include: (i) Most nodes have a low degree, and the network exhibits a heavy-tailed degree distribution. (ii) A small number of nodes have a very high degree, making the presence of so-called \textit{hubs} a defining feature. (iii) There is no characteristic scale for node connectivity, which gives rise to the term "scale-free". (iv) In comparison to random graphs, scale-free networks are known as more robust to random failures because highly connected hubs are rarely selected at random, although they remain comparatively vulnerable to targeted attacks\citep{Rohrer2019}. Similar have also been observed across a broad collection of empirical scale-free networks when compared with preferential-attachment models\citep{Adami2026}.

Several point estimates for power-law curves have appeared in the LN literature. For instance, in \citep{Rohrer2019}, the reported estimate of $|\alpha|$ is 2.18, while in \citep{Mahdizadeh2023} it is 2.28. These findings provide fragmentary support for the conclusion that the LN exhibits power-law characteristics over time. In this study, we aimed to extensively verify the scale-free nature and power-law degree distribution of the LN over a broader time frame. To this end, we adopted the methodology proposed in \citep{Rohrer2019}, namely, fitting a power-law curve to the empirical degree distributions of the LN snapshots.

\begin{figure*}[t]
\centering
\hspace{-0.5cm}
\includegraphics[scale=0.15, center]{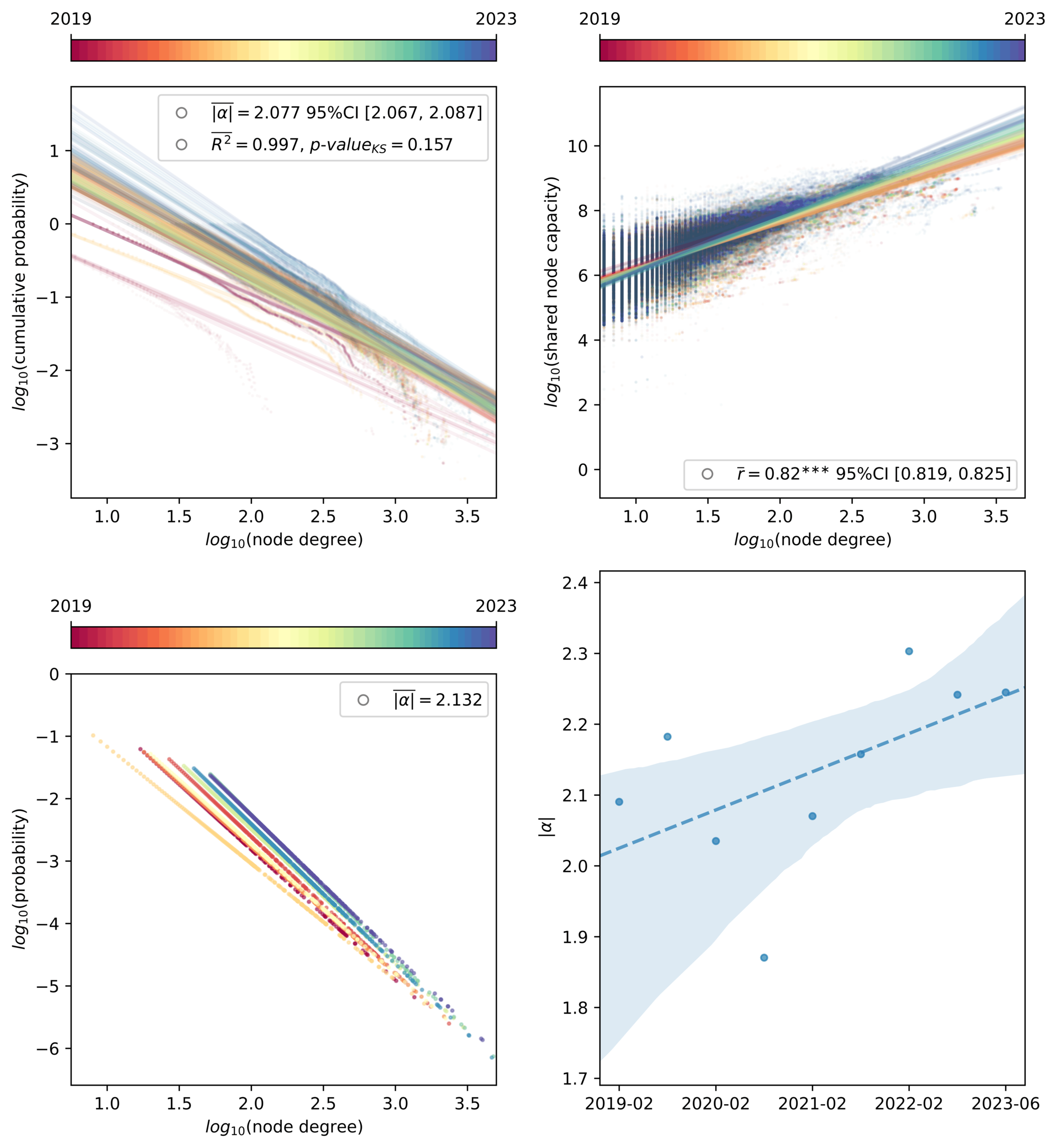}
\caption{Degree distribution of the network and the associated shared node capacity over time.}\label{fig03}
\begin{tablenotes}[]\footnotesize\item[]
Note. The upper-left panel presents a log-log plot of the cumulative degree distribution of the network (directed representation). Points correspond to the empirical distributions for individual snapshots, and solid lines denote the fitted power-law models\citep{pl}. Colours indicate the snapshot date (see the colour bar above). The upper-right panel shows the log–log relationship between node degree and shared node capacity, as defined in Eq.~\eqref{eq8}. Points represent the empirical data, while solid lines indicate the corresponding fitted models. The average Spearman's correlation coefficient ($r$) and power-law exponent ($|\alpha|$) is reported together with bootstrapped ($n = 100$) 95\% confidence intervals. The bottom-left panel displays nine selected snapshots, spaced at approximately six-month intervals, to illustrate the temporal shift in the degree distribution; the average estimated $|\alpha|$ across these snapshots is also reported. The bottom-right panel shows the estimated values of $|\alpha|$ for the same snapshots, together with their linear fit and 95\% confidence intervals.
\end{tablenotes}
\end{figure*}

\textit{Figure~\ref{fig03}} (upper-left panel) presents a graphical representation of the empirical degree distributions across all snapshots together with the corresponding fitted power-law curves. It jointly visualizes the heavy-tailed degree distribution and also its strong association with shared node capacity. The log–log linearity and stable exponent estimates across snapshots suggest persistent hub dominance, while the tight coupling between degree and capacity highlights the economic relevance of topological centrality.

The curves exhibit a good fit using the discrete power-law estimation package\citep{pl}. The average estimated power-law exponent $|\alpha|$ equals 2.077, while the bootstrapped coefficient of determination is 0.997. The Kolmogorov–Smirnov test yields an insignificant statistic ($p$-value = 0.157), indicating that the power-law model provides an adequate fit to the data. Although the power-law model provides a statistically plausible fit over most of the observed snapshots, we refrain from claiming strict scale-free behaviour in the sense of Clauset et al.\citep{Clauset}. Rather, the LN exhibits persistent heavy-tailed degree distributions consistent with hub dominance. All fitting statistics, likelihood-ratios, and optimal values for \textit{xmin} are reported in the supplementary repository. 

The strong positive correlation (0.82) between node degree and shared node capacity implies that structural prominence translates directly into economic influence. High-degree nodes not only connect widely but are also able control a disproportionately large share of routing liquidity, see \textit{Figure~\ref{fig03}} (upper-right panel). Thus, in the following analysis, all statements about degree distribution can be applied to shared node capacity distribution, unless otherwise specified.

The bottom-left panel on the \textit{Figure~\ref{fig03}} displays nine selected snapshots, spaced at approximately six-month intervals, to better illustrate the temporal shift in the degree distribution. The average estimated $|\alpha|$ across these snapshots is 2.13. The bottom-right panel shows the exact $|\alpha|$ estimates for the same snapshots, together with a linear fit, highlighting an upward trend in the power-law exponent.

\begin{figure*}[t]
\includegraphics[scale=0.55, center]{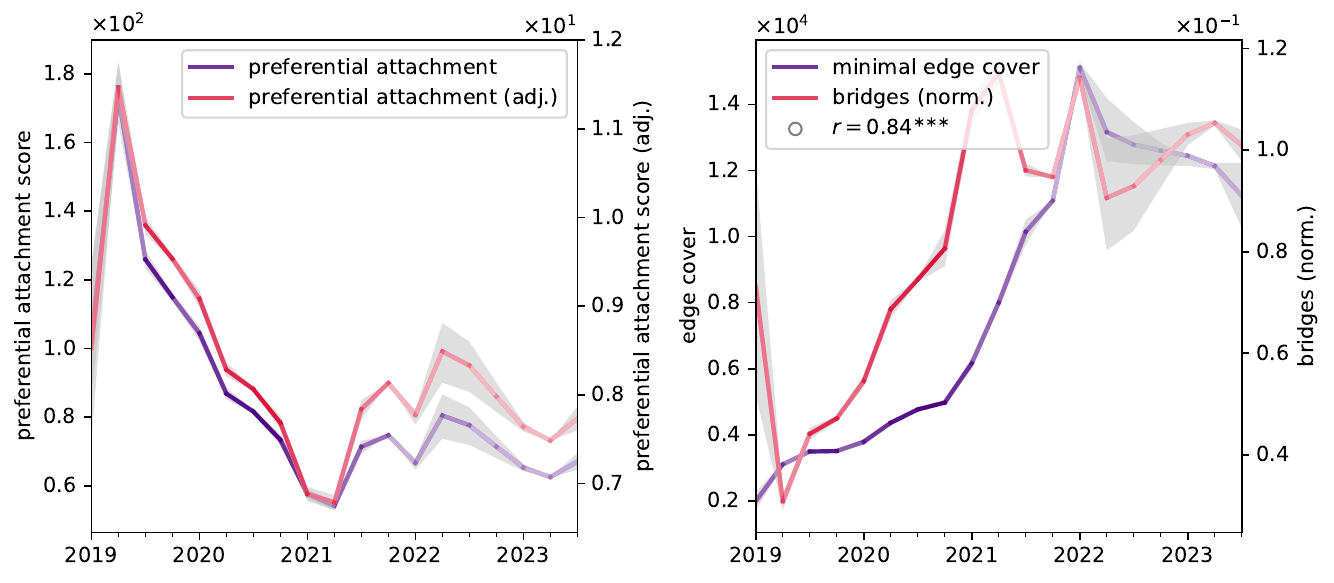}
\caption{The average preferential attachment score, bridges and minimal edge cover over time.}\label{fig04}
\begin{tablenotes}[]\footnotesize\item[]
Note. Quarterly averaged estimates (solid lines) are presented together with bootstrapped ($n = 100$) 95\% confidence intervals (shaded areas). Line colour intensity reflects the normalized data coverage for each time period based on the available snapshots. Significance levels for Spearman's correlation $p$-value: *** $< 0.001$, ** $< 0.01$, * $< 0.05$.
\end{tablenotes}
\end{figure*}

As this type of network structure emerges naturally from a process known as \textit{preferential attachment}\citep{Nowell}, it can be illustrated by the principle of "the rich get richer", when nodes with high degrees are more likely to form new connections. Further illustrating this, the average preferential attachment score shows a declining trend over time in the LN, see \textit{Figure~\ref{fig04}}. A decline of comparable magnitude is also observed in the average node degree (\textit{Figure~\ref{fig02}}). When examining the average preferential attachment score normalized by the average node degree, we observe that a slight tendency toward reduced preferential attachment has persisted over time. This suggests that while preferential attachment may have shaped the early LN topology, later evolution appears dominated by node retention consistent with reduced marginal hub intensification.

While preferential attachment explains how new edges are formed, it does not reveal how robust the resulting structure is. Therefore, we next analyse the LN's connectivity and resilience, focusing on how well the network withstands edge removals or node failures.

\subsubsection{Connectivity and resilience}

The foundational metric for analysing network connectivity is \textit{the number of bridges} (or cut-edges) -- edges whose removal increases the number of connected components within the graph. Intuitively, a bridge represents a single point of failure: removing such an edge disconnects part of the network. For example, if a peripheral LN node is connected to the network through only one payment channel, that channel forms a bridge.

A higher number of bridges typically indicates a more technically fragile network structure, characterized by numerous single points of failure. As illustrated in \textit{Figure~\ref{fig04}}, the number of bridges rises noticeably after 2020, suggesting that the LN might become increasingly technically vulnerable to targeted node or channel failures despite its overall growth. Note that the number of bridges here is normalised by the number of channels in order to observe structural changes without the direct influence of network growth.

A related concept is the \textit{minimal edge cover}, defined as a set of edges such that every node is incident to at least one edge. A smaller edge cover suggests a sparse yet efficient configuration that guarantees basic node reachability. As a simple illustration, in a star topology every peripheral node is covered by its single incident edge, so the minimum edge cover contains all star edges. In more densely connected networks, fewer strategically selected edges are sufficient to cover all nodes.

In the context of the LN, the size of the minimal edge cover increases over time (see \textit{Figure~\ref{fig04}}), indicating growing connectivity demands, which align with the observed increase in the number of bridges. As expected, the size of the minimal edge cover and the number of bridges are highly correlated ($>$ 0.8). The parallel increase in minimal edge cover and bridge count suggests that new connectivity is not primarily reinforcing redundancy but rather expanding coverage with structurally critical links. Importantly, an increase in bridges does not imply immediate routing failure, as many bridges correspond to high-capacity or long-lived channels that remain operational over extended periods (this will be shown later).

\begin{figure*}[t]
\centering
\hspace{-1.2cm}
\includegraphics[scale=0.55, center]{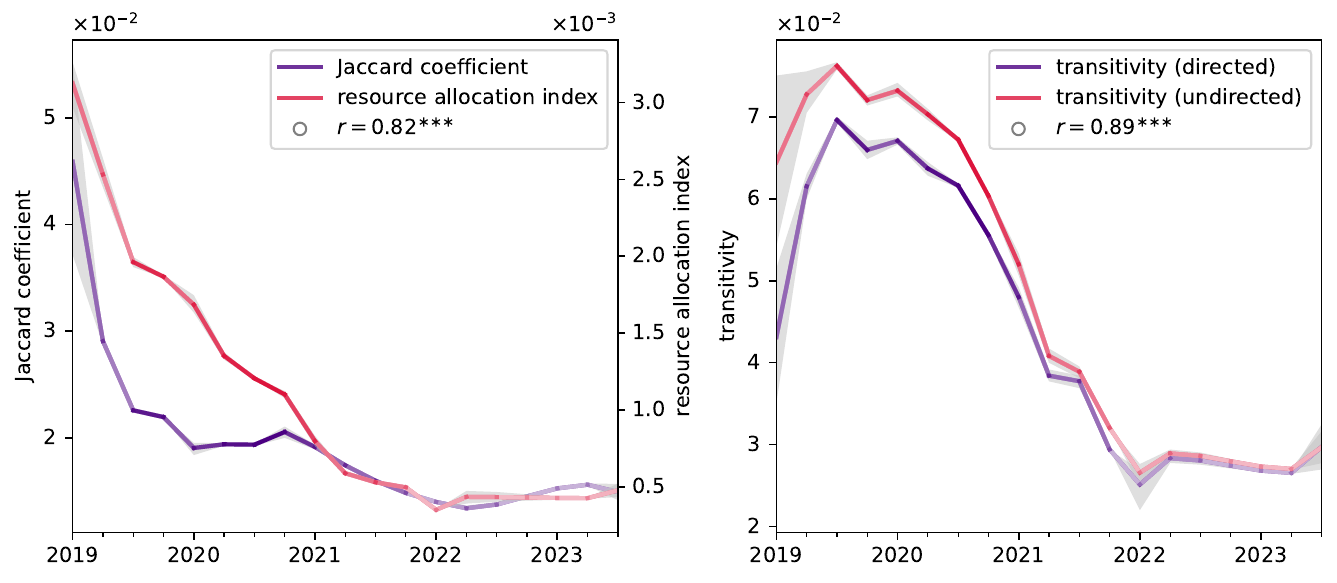}
\caption{The average resource allocation index, Jaccard coefficient and transitivity over time.}\label{fig05}
\begin{tablenotes}[]\footnotesize\item[]
Note. Quarterly averaged estimates (solid lines) are presented together with bootstrapped ($n = 100$) 95\% confidence intervals (shaded areas). Line colour intensity reflects the normalized data coverage for each time period based on the available snapshots. Significance levels for Spearman's correlation $p$-value: *** $< 0.001$, ** $< 0.01$, * $< 0.05$.
\end{tablenotes}
\end{figure*}

Transitivity, or the global clustering coefficient, measures the probability that two neighbours of a node are themselves connected. Higher transitivity indicates greater local redundancy, thereby enhancing fault tolerance. Conversely, lower transitivity is associated with more loosely connected subgraphs, reducing technical robustness. This latter pattern in the LN is illustrated in \textit{Figure~\ref{fig05}}. Notably, the result holds for both directed and undirected graph representations, which exhibit a strong correlation of 0.89.

\begin{figure*}[t]
\centering
\hspace{-1.2cm}
\includegraphics[scale=0.5, center]{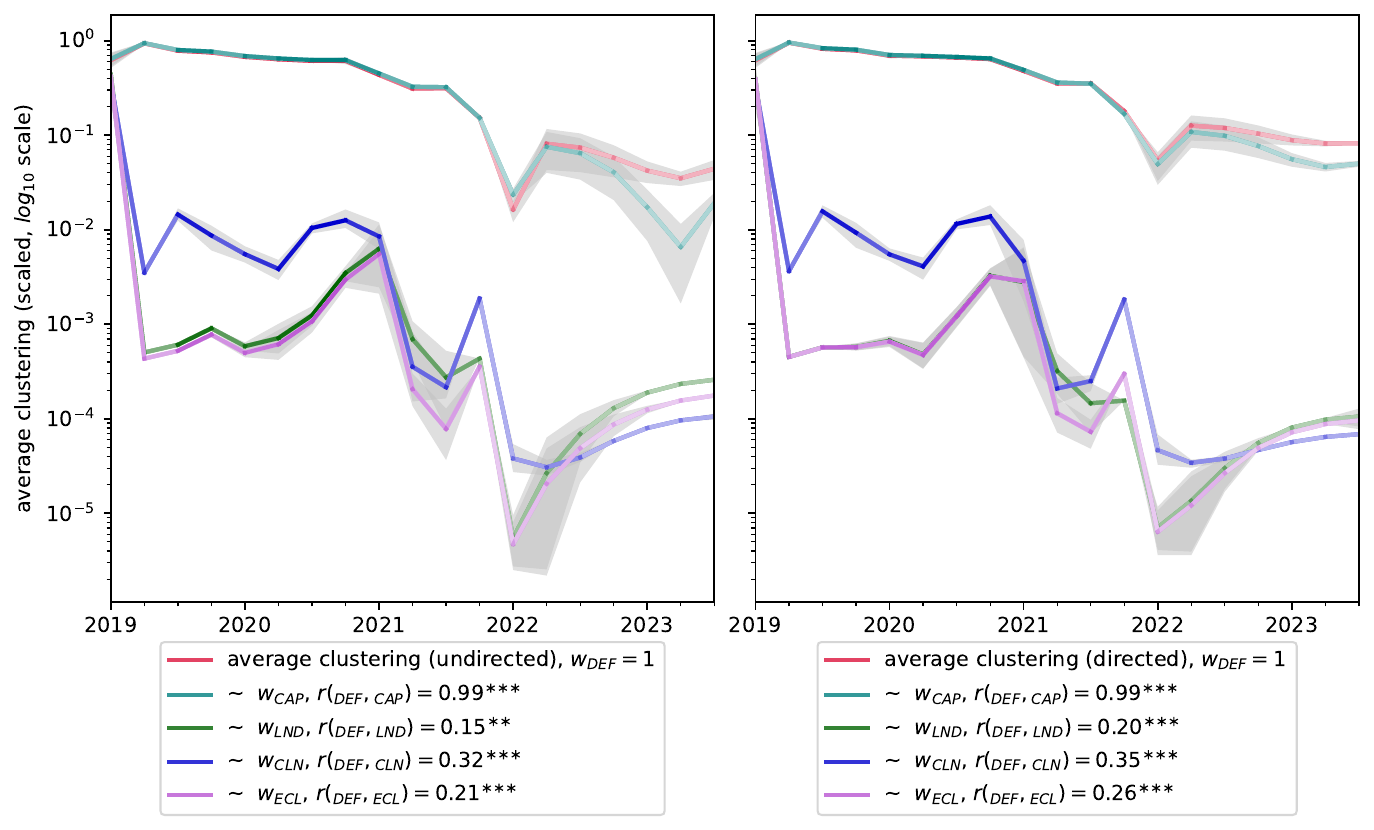}
\caption{The average clustering coefficient over time.}\label{fig06}
\begin{tablenotes}[]\footnotesize\item[]
Note. Quarterly averaged estimates (solid lines) are presented together with bootstrapped ($n = 100$) 95\% confidence intervals (shaded areas). Line colour intensity reflects the normalized data coverage for each time period based on the available snapshots.
All calculations are performed for both undirected (left panel) and directed (right panel) graph representations, considering the unweighted case ($w_{DEF} = 1$), the capacity-weighted specification ($w_{CAP}$), and realistic pathfinding strategies ($w_{LND}$, $w_{CLN}$, $w_{ECL}$). Spearman's correlation coefficients are reported relative to the simple unweighted case. Significance levels for Spearman’s correlation $p$-values are denoted as follows: *** $< 0.001$, ** $< 0.01$, * $< 0.05$.
\end{tablenotes}
\end{figure*}

Similarly, the average clustering coefficient, defined as the mean of the local clustering coefficients across all nodes, provides insight into the network’s local structure. In the case of the LN, declining average clustering values formally indicate increasing sparsity and reduced technical resilience of local connections (see \textit{Figure~\ref{fig06}}).

We compute this metric using both directed and undirected graph representations and account for realistic pathfinding strategies ($w_{CAP}$, $w_{LND}$, $w_{CLN}$, $w_{ECL}$, see \textit{Figure~\ref{fig06}}). Although the absolute magnitudes differ due to differences in weighting schemes, the overall trend can be meaningfully compared across specifications. In all cases, the average clustering coefficient exhibits a declining trend, despite slight variations across pathfinding strategies. 

Notably, the capacity-weighted measure is highly correlated with the unweighted measure and all weighting schemes perform consistently across both directed and undirected representations. Despite differences in absolute magnitude, all specifications exhibit a consistent declining trend, indicating that sparsification is robust to modelling assumptions about routing costs. The strong correlation between unweighted and capacity-weighted clustering indicates that structural cohesion is largely topology-driven rather than capacity-driven. 

Overall, this analysis indicates several structural characteristics that may increase susceptibility to targeted disruptions, although robustness under adversarial failures was not directly evaluated. For instance, the growing number of bridges points to a more fragile topology, while low clustering metrics may suggest insufficient local redundancy. These indicators combined highlight the need for strategic improvements in the network's design to enhance both global structural robustness and local channel fault tolerance.

Beyond structural robustness, it is also essential to understand how the LN's topology affects information flow and operational efficiency over time. The following subsection examines these dynamic aspects of network function.

\subsubsection{Function and dynamics}

To analyse the function and dynamics of the LN, it is important to understand how efficiently information flows through the network, which nodes are critical to that flow, and how robust or vulnerable the structure is to changes. For this analysis, we report three the most important measures: \textit{global efficiency}, \textit{effective size / Burt's effective size}, and \textit{information centrality}.

\begin{figure*}[t]
\hspace{-1.2cm}
\includegraphics[scale=0.6, center]{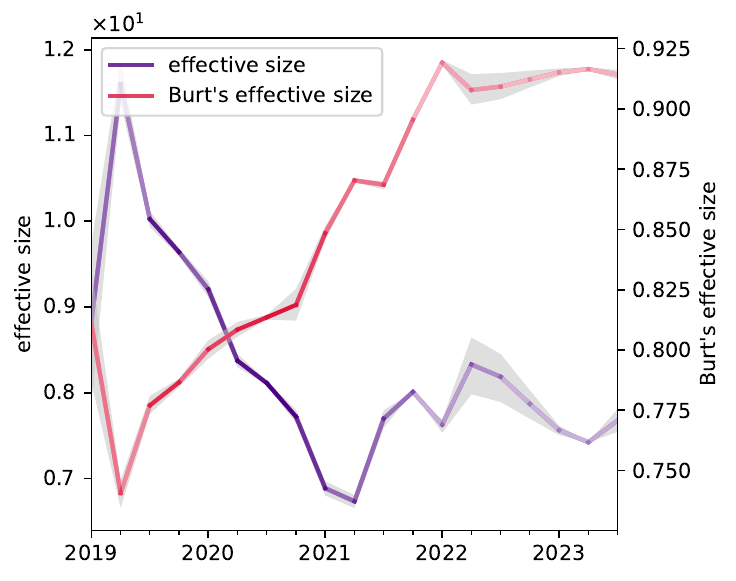}
\caption{Effective size and Burt's effective size over time.}\label{fig07}
\begin{tablenotes}[]\footnotesize\item[]
Note. Quarterly averaged estimates (solid lines) are presented together with bootstrapped ($n = 100$) 95\% confidence intervals (shaded areas). Line colour intensity reflects the normalized data coverage for each time period based on the available snapshots. Significance levels for Spearman's correlation $p$-value: *** $< 0.001$, ** $< 0.01$, * $< 0.05$.
\end{tablenotes}
\end{figure*}

As shown above, both the average degree and network density decline over time, indicating that nodes are forming relatively fewer connections. As the network becomes sparser, the focal nodes are expected to exhibit improved local efficiency, reflected in an increase in Burt's effective size (see \textit{Figure~\ref{fig07}}). Note that Burt’s effective size differs from the conventional effective size measure in that it normalizes the value by the node’s degree. This suggests that while the number of connections may have decreased, the remaining ties are less redundant and more structurally valuable. However, the simultaneous decline in general effective size, information centrality, and global efficiency reveals a more nuanced picture. The increase in normalized effective size reflects decreasing redundancy in ego networks, whereas the decline in global efficiency captures longer average path lengths at the system level. These trends are not contradictory but reflect multi-scale restructuring.

\begin{figure*}[t]
\hspace{-1.2cm}
\includegraphics[scale=0.55, center]{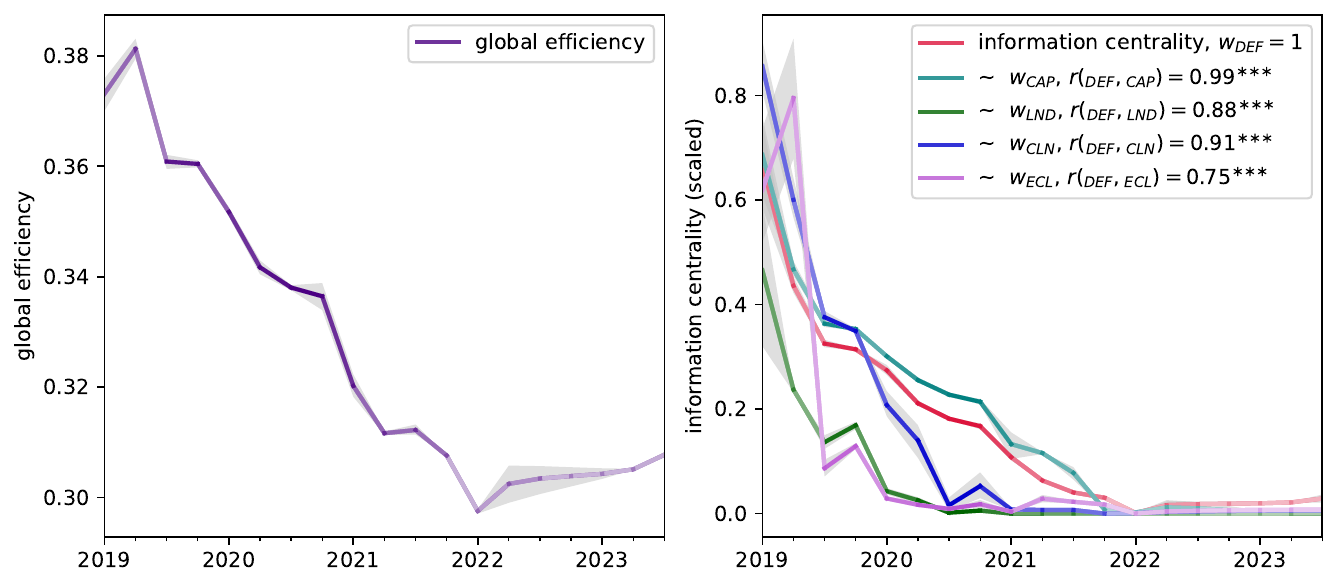}
\caption{Global efficiency and information centrality over time.}\label{fig08}
\begin{tablenotes}[]\footnotesize\item[]
Note. Quarterly averaged estimates (solid lines) are presented together with bootstrapped ($n = 100$) 95\% confidence intervals (shaded areas). Line colour intensity reflects the normalized data coverage for each time period based on the available snapshots.
Information centrality calculations are performed for undirected graph representations, considering the unweighted case ($w_{DEF} = 1$), the capacity-weighted specification ($w_{CAP}$), and realistic pathfinding strategies ($w_{LND}$, $w_{CLN}$, $w_{ECL}$). Spearman's correlation coefficients are reported relative to the simple unweighted case. Significance levels for Spearman's correlation $p$-values are denoted as follows: *** $< 0.001$, ** $< 0.01$, * $< 0.05$.
\end{tablenotes}
\end{figure*}

Although a node may retain a strategic position within its immediate neighbourhood, its overall technical access to the broader network may be diminishing. The observed decline in information centrality over the analysed time frame suggests a reduced role in facilitating communication between other nodes (see \textit{Figure~\ref{fig08}}), while the decrease in global efficiency indicates that the network as a whole may be becoming less effective at transmitting information rapidly and extensively. Notably, all weighted measures of information centrality are highly correlated over time (greater than 0.8), and even the simple unweighted representations capture the overall pattern well.

Taken together, these trends point to a shift toward a less cohesive network -- one that may be increasingly constrained in supporting efficient large-scale information flow, under the theoretical definition of information flow for this measure\citep{Brandes2005CentralityMB}. In this setting, a node can preserve local structural importance while simultaneously losing broader influence and reach. Thus, consistent with the connectivity and resilience analysis above, although a node's remaining ties may be non-redundant and structurally valuable, its overall influence and accessibility may be declining as the LN becomes more fragmented.

Given the LN's primary function as a payment network, full reachability of all nodes may not be essential for transaction routing. However, from the perspective of propagating gossip messages that ensure network consistency, these findings may carry particular significance for LN developers.

\subsubsection{Emergent patterns}

This analysis provides a more detailed understanding of how the network patterns evolve over time. While it was previously shown that the network is growing in size, it is seemingly not becoming more cohesive. In fact, metrics such as the \textit{resource allocation index} and the \textit{Jaccard coefficient} suggest that new connections are more likely to be cross-cutting or bridging previously isolated regions, rather than reinforcing existing clusters (see \textit{Figure~\ref{fig05}}).

\begin{figure*}[t]
\hspace{-1.7cm}
\includegraphics[scale=0.5, center]{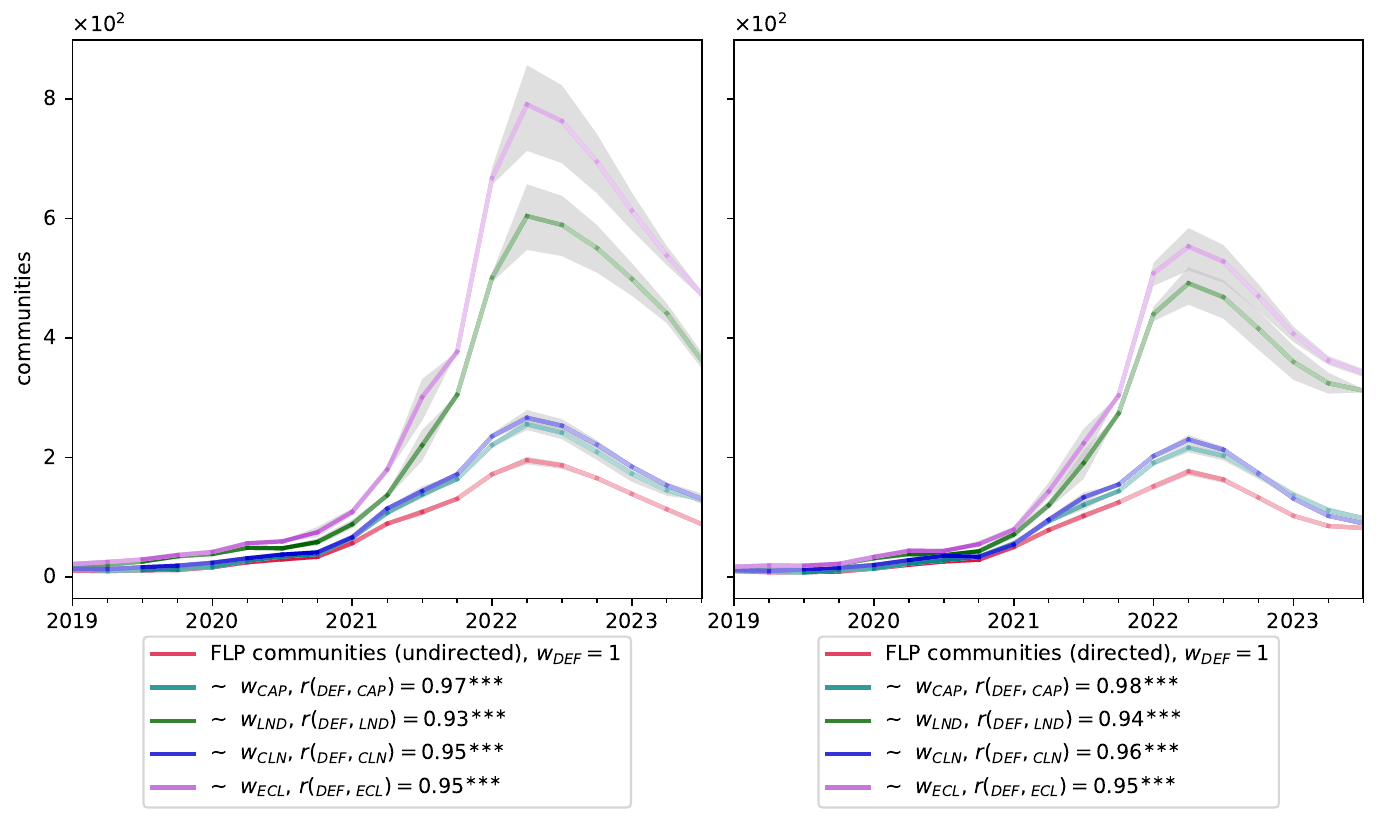}
\hspace{-1.7cm}
\includegraphics[scale=0.5, center]{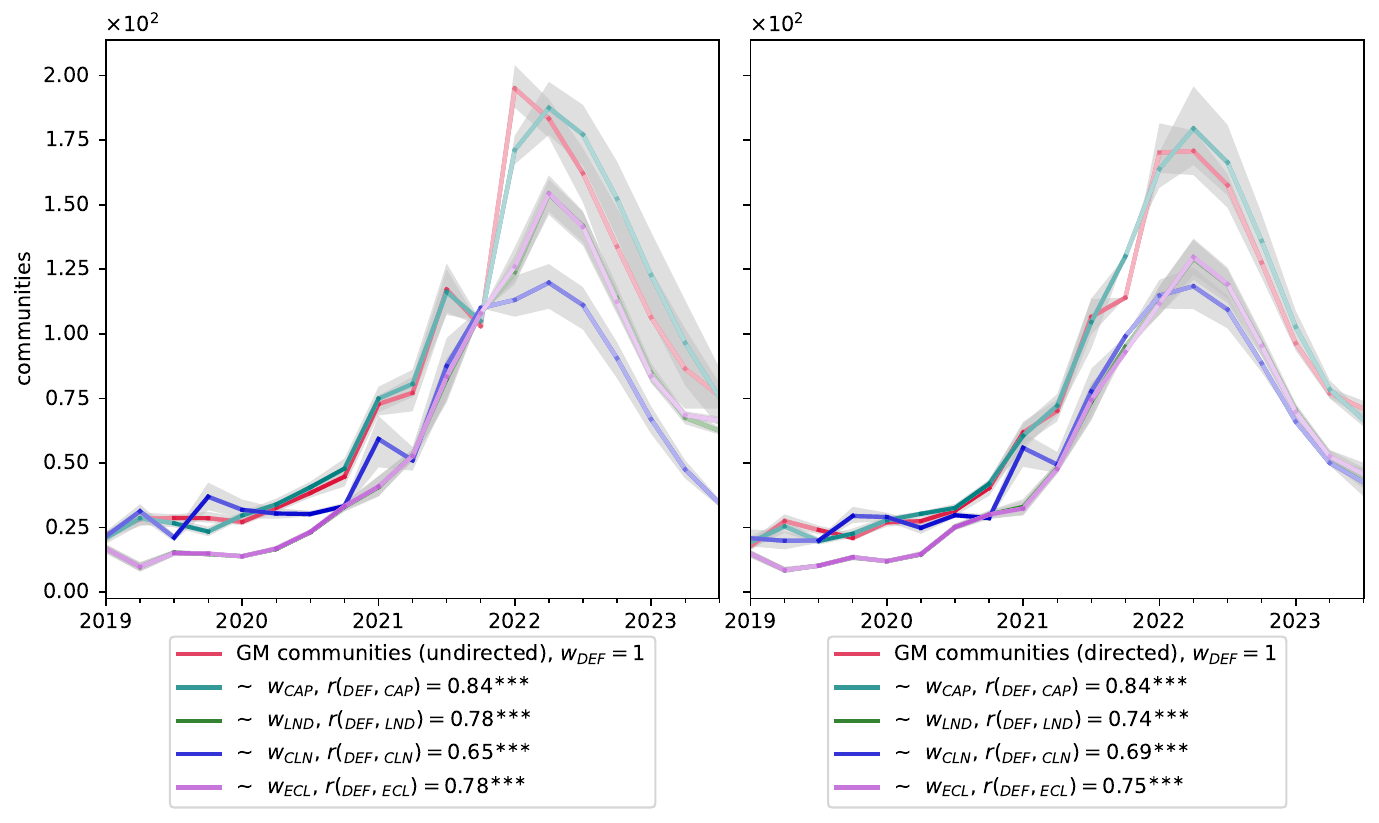}
\caption{Network FLP communities (upper panel) and GM communities (lower panel) over time.}\label{fig09}
\begin{tablenotes}[]\footnotesize\item[]
Note. Quarterly averaged estimates (solid lines) are presented together with bootstrapped ($n = 100$) 95\% confidence intervals (shaded areas). Line colour intensity reflects the normalized data coverage for each time period based on the available snapshots.
All calculations are performed for both undirected (left panel) and directed (right panel) graph representations, considering the unweighted case ($w_{DEF} = 1$), the capacity-weighted specification ($w_{CAP}$), and realistic pathfinding strategies ($w_{LND}$, $w_{CLN}$, $w_{ECL}$). Spearman's correlation coefficients are reported relative to the simple unweighted case. Significance levels for Spearman’s correlation $p$-values are denoted as follows: *** $< 0.001$, ** $< 0.01$, * $< 0.05$.
\end{tablenotes}
\end{figure*}

The observed decline in both the resource allocation index and the Jaccard coefficient, combined with an increase in the number of detected communities (as estimated using the fast label propagation (FLP\citep{Traag2023}) and the greedy modularity (GM\citep{ClausetNewmanMoore2004}) algorithms), points to a network that is becoming increasingly fragmented, see \textit{Figure~\ref{fig09}}. Notably, all community measures are highly correlated across graph representations ($>$ 0.9).

A comparably lower resource allocation index indicates that nodes are sharing fewer common neighbours, which reduces the potential for efficient information or resource flow based on local network structures. Similarly, the declining Jaccard coefficient reflects a decrease in neighbourhood overlap between connected nodes, suggesting that ties are becoming less cohesive and more diverse. Meanwhile, the growing number of communities implies that the network is splitting into a larger number of smaller, tightly-knit groups that are internally cohesive but more isolated from one another. 

Together, these trends indicate decreasing structural cohesion and increasing modular organization of the observable topology. The simultaneous decline in neighbourhood overlap and increase in the number of detected communities is consistent with progressively stronger community structure. However, because community-detection algorithms are known to exhibit scale-dependent behaviour as network size increases\citep{Leskovec2010}, these findings should be interpreted as evidence of increasing modularization rather than definitive proof of topological fragmentation. Confirming whether the observed increase exceeds that expected from network growth alone would require comparisons against appropriate degree-preserving null models, which we leave for future work.

To assess whether this increasing modular organization affects the network's transactional stability and established routing behaviour, we next turn to a detailed analysis of temporal–topological and operational stability.

\subsection{Stability and centrality analysis}
\label{sec_results_sub2}

In this section, we analyse the \textit{temporal–topological stability} of the LN from two complementary perspectives: (i) the technical and statistical retention of nodes and channels across successive snapshots, and (ii) capacity-based and operational stability, reflecting the observed cost-based feasibility of nodes and channels within the evolving network. Particular attention is also given to network decentralization and inequality, which also has its implications for overall transactional stability.

It is also important here to distinguish between (i) base-level technical and distributional stability, (ii) structural and topological cohesion at the mesoscopic level (communities, clustering etc.), and (iii) operational channel feasibility and observed routing-level stability under weighted pathfinding (see, \textit{Table~\ref{tab:stability}}). The following results show that while mesoscopic cohesion declines (see the network analysis above), distributional and routing-level stability remain high. This indicates that increasing modular organization and possible structural fragmentation do not yet translate into functional and economical instability.

\begin{table*}[!htbp]
\caption{Operationalising temporal–topological stability.}
\label{tab:stability}
\centering
\small
\begin{tabular}{|p{.3\textwidth}|p{.3\textwidth}|p{.3\textwidth}|}
\hline
\textbf{Stability type} & \textbf{Key measures} & \textbf{Focus} 
\\\hline
\textit{Structural (topological)} & Density, clustering, communities etc. & Observable graph cohesion
\\\hline
\textit{Technical (distributional)} & Node/channel retention, degree distribution stability & Observable technical persistence
\\\hline
\textit{Operational (routing-level)} & Weighted channel retention, shared node capacity distribution stability & Observable routing feasibility
\\\hline
\end{tabular}
\end{table*}

From the first perspective, we examine the LN's degree distribution as a fundamental topological characteristic and conduct statistical tests to assess its temporal similarity. In particular, we apply the Kolmogorov-Smirnov (KS) test, which evaluates whether two successive snapshots are drawn from the same underlying degree distribution. Under the null hypothesis, the distributions are considered identical; a $p$-value below 0.05 leads to its rejection. 

\begin{figure*}[t]
\hspace{-3cm}
\includegraphics[scale=0.55, center]{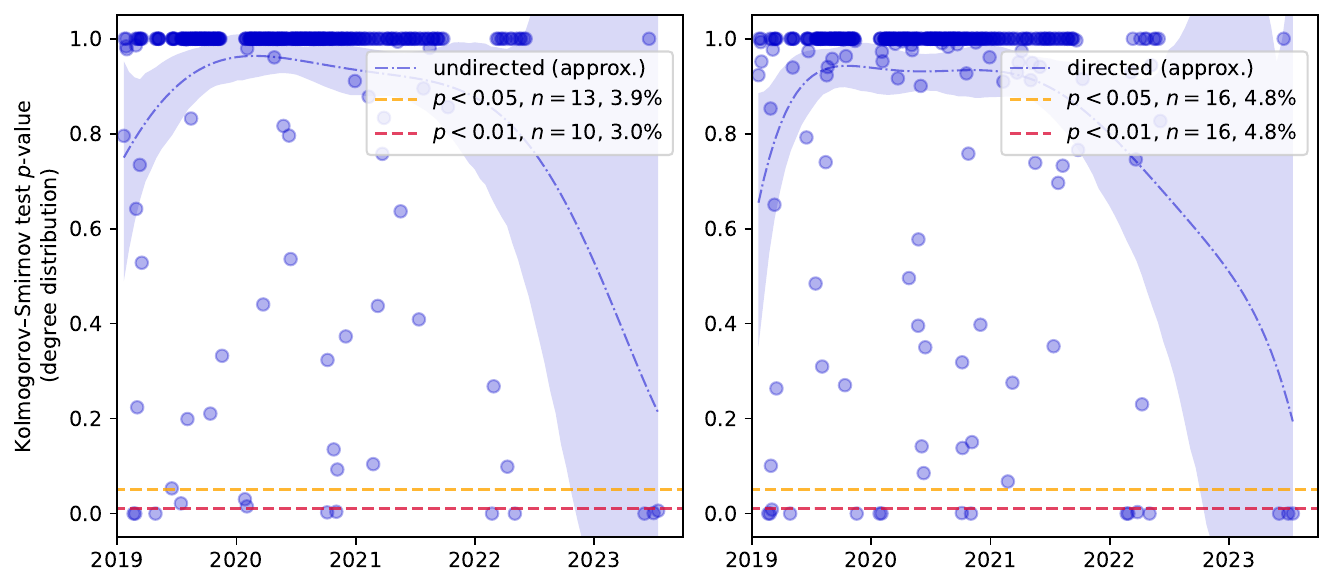}
\hspace{-3cm}
\includegraphics[scale=0.55, center]{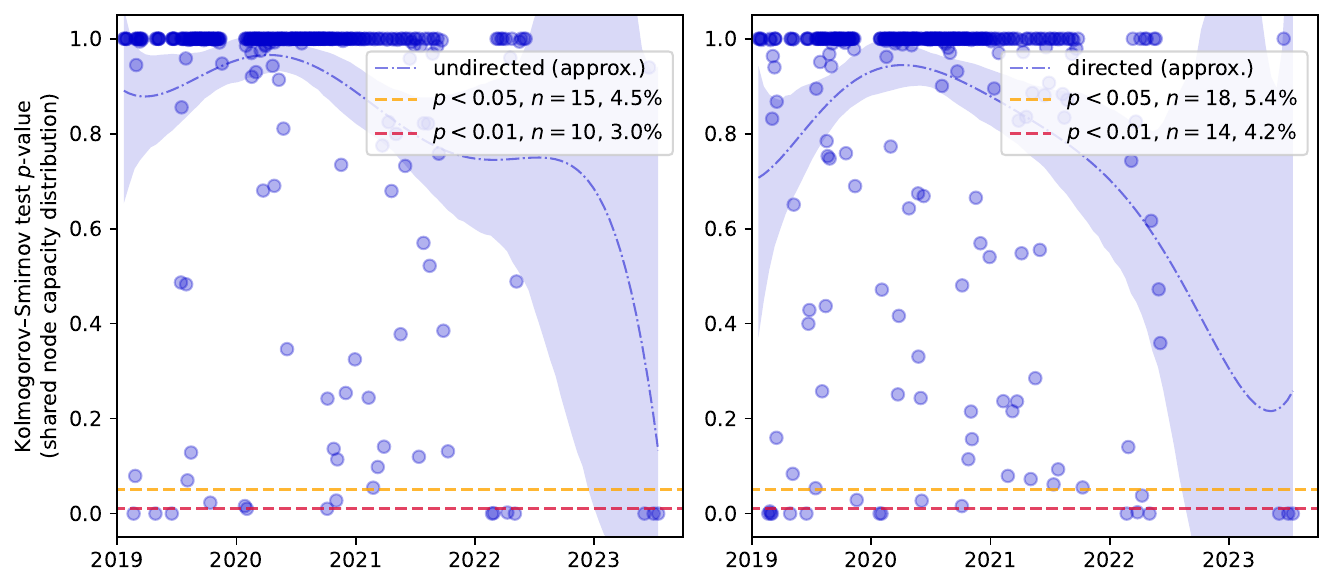}
\caption{The Kolmogorov-Smirnov test results for the node degree distribution (upper panel) and the shared node capacity distribution (lower panel).}\label{fig10}
\begin{tablenotes}[]\footnotesize\item[]
Note. Each point represents a comparison between two successive snapshots. Calculations are performed for both undirected (left panel) and directed (right panel) graph representations. The number of comparisons in which the null hypothesis for KS test is likely rejected is reported both as an absolute count ($n$) and as a percentage.
\end{tablenotes}
\end{figure*}

\textit{Figure~\ref{fig10}} illustrates the temporal distribution of $p$-values obtained from the KS tests. As shown, fewer than 5\% of transitions between successive snapshots are statistically significant (i.e., $p < 0.05$). These significant changes typically occur near large data gaps, which can be observed as regions with sparse dot concentration along the time axis (see \textit{Figure~\ref{fig10}}). This empirical observation holds for directed and undirected graph representations and for both the degree distribution (upper panel) and the shared capacity distribution (lower panel), the latter serving as a proxy for the economic stability of payment flows.

To complement this analysis, we compute the Wasserstein distance\citep{Ramdas} (WD) between corresponding degree distributions. This metric measures the minimum "cost" required to transform one distribution into another, with smaller values indicating greater similarity. 

\begin{figure*}[t]
\hspace{-3cm}
\includegraphics[scale=0.55, center]{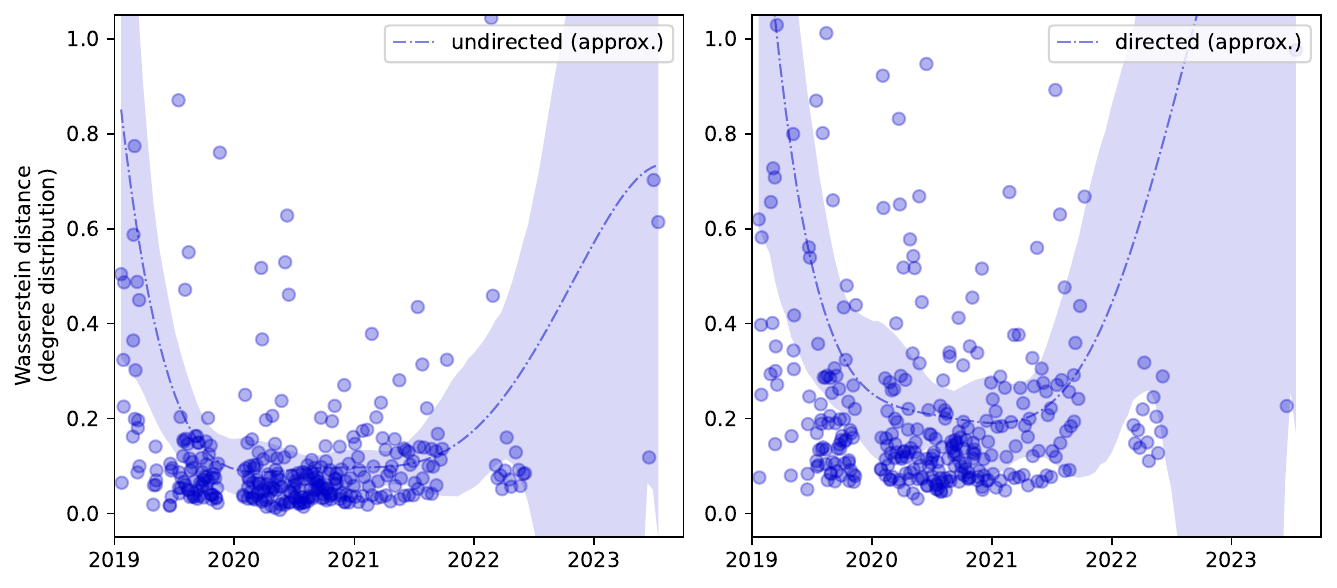}
\hspace{-3cm}
\includegraphics[scale=0.55, center]{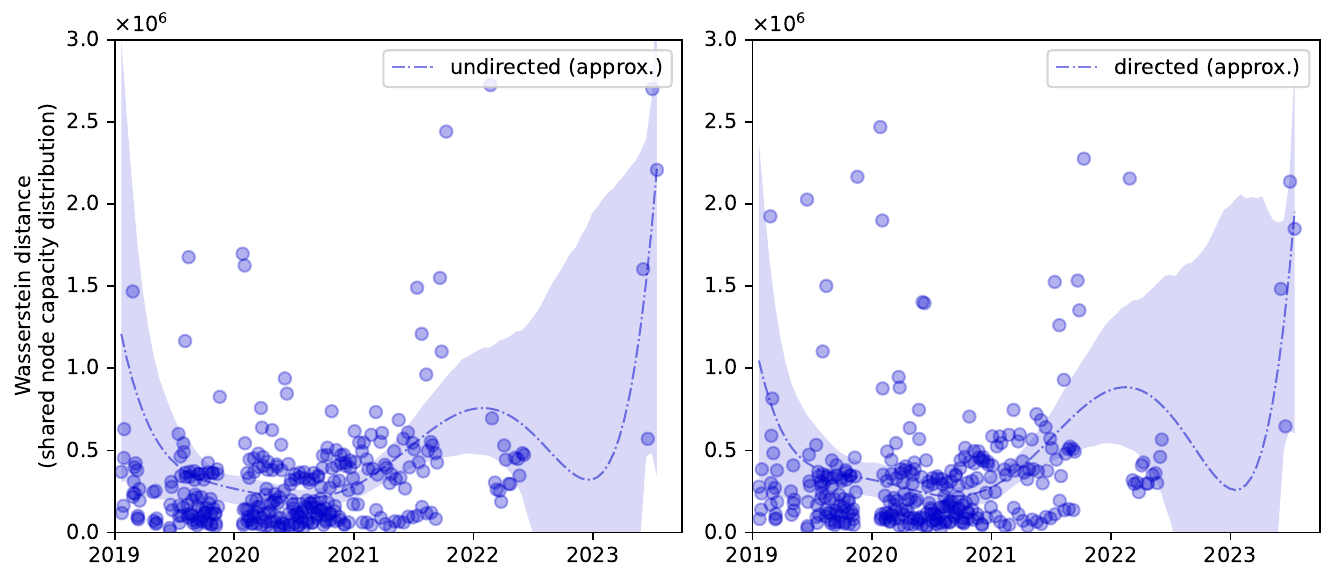}
\caption{The Wasserstein distance results for the node degree distribution (upper panel) and the shared node capacity distribution (lower panel).}\label{fig11}
\begin{tablenotes}[]\footnotesize\item[]
Note. Each point represents a comparison between two successive snapshots. Calculations are performed for both undirected (left panel) and directed (right panel) graph representations.
\end{tablenotes}
\end{figure*}

The \textit{Figure~\ref{fig11}} shows the WD between corresponding degree distributions (upper panel) and the shared capacity distributions (lower panel) of successive snapshots. As observed, the trend of WDs closely mirrors the KS $p$-value behaviour, with notable deviations primarily occurring also around documented data gaps. The convergence between KS and Wasserstein diagnostics confirms that distributional stability is not a statistical artefact.

On average, the absolute distance remains small: 0.2 (undirected) and 0.4 (directed) for the degree distribution, and $3.8 \times 10^{5}$ SAT for the shared capacity distribution (approx. 0.0038 BTC). Note that, as of 25 February 2026, the estimated total LN capacity is 2,572.58 BTC (see https://1ml.com/statistics). Thus, while the WD is expressed in absolute capacity units (SAT), its magnitude remains negligible relative to the evolving total capacity scale, indicating limited distributional shift.

Thus, \textit{Figures~\ref{fig10}} and \textit{\ref{fig11}} provide complementary statistical evidence for temporal distributional stability. While occasional deviations coincide with data gaps, both KS statistics and WDs indicate that successive degree and capacity distributions remain similar across most of the observation period. The joint interpretation of KS (sensitive to maximum deviation) and WD (sensitive to global mass shifts) ensures robustness against metric-specific and parametric artefacts.

The second perspective builds upon the statistical and technical retention results and evaluates whether the network preserves its efficiency under realistic payment pathfinding strategies, specified through the weighting schemes in Eqs. \eqref{eq3}--\eqref{eq5}. This computationally intensive analysis also serves to determine whether the weighted schemes significantly differ from the simplified approaches commonly adopted in LN empirical studies.

\begin{figure*}[t]
\centering
\hspace{-2.5cm}
\includegraphics[scale=0.55, center]{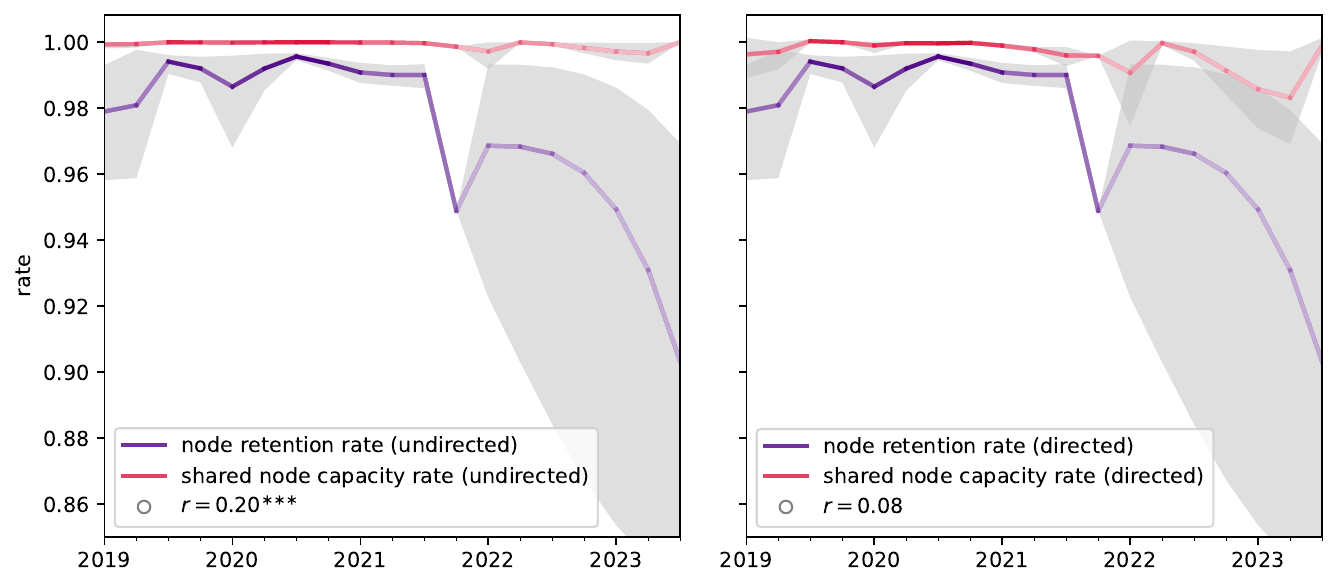}
\caption{The node retention rate and the shared node capacity rate over time.}\label{fig12}
\begin{tablenotes}[]\footnotesize\item[]
Note. Quarterly averaged estimates (solid lines) are presented together with bootstrapped ($n = 100$) 95\% confidence intervals (shaded areas). Line colour intensity reflects the normalized data coverage for each time period based on the available snapshots. The node retention rate $(I_{node})$ and the shared node capacity rate $(I_{node}^{capacity})$ are calculated as defined in Eqs. \eqref{eq6} and \eqref{eq9}, respectively. Significance levels for Spearman’s correlation $p$-values are denoted as follows: *** $< 0.001$, ** $< 0.01$, * $< 0.05$.
\end{tablenotes}
\end{figure*}

\textit{Figure~\ref{fig12}} presents our proxy measures for observed operational stability: the \textit{node retention rate} and the \textit{shared node capacity rate} of retained nodes. The node retention rate remains higher than 0.9 throughout nearly the entire observation period, while the shared capacity rate stays particularly close to 1.0. This indicates that the vast majority of nodes persist between successive snapshots and that almost the same share of the total network's capacity retained in these stable nodes.

As expected, node retention rates are negatively correlated with the time in days elapsed between successive snapshots: the longer the interval, the lower the retention rate. The Spearman's correlation coefficient for node retention rate across the entire time frame is approx. $-0.8$ ($p < 0.001$, for both directed and undirected representations). In contrast, the shared node capacity rate is less sensitive to time, with a weaker correlation of approx. $-0.3$ ($p < 0.001$, for both directed and undirected representations).

These findings suggest that the observed operational state of the LN remains highly stable over time. Even the direct comparison between the earliest available snapshot (20-01-2019) and the most recent one (16-07-2023) reveals substantial overlap: approx. $31$\% of nodes and $30$\% of shared capacity are common to both snapshots. This long-term persistence is further supported by available statistics, which show that the average node age as of 25 February 2026 is 1,331.2 days (approx. 3.6 years, see https://1ml.com/statistics).

\begin{figure*}[t]
\centering
\hspace{-2.5cm}
\includegraphics[scale=0.5, center]{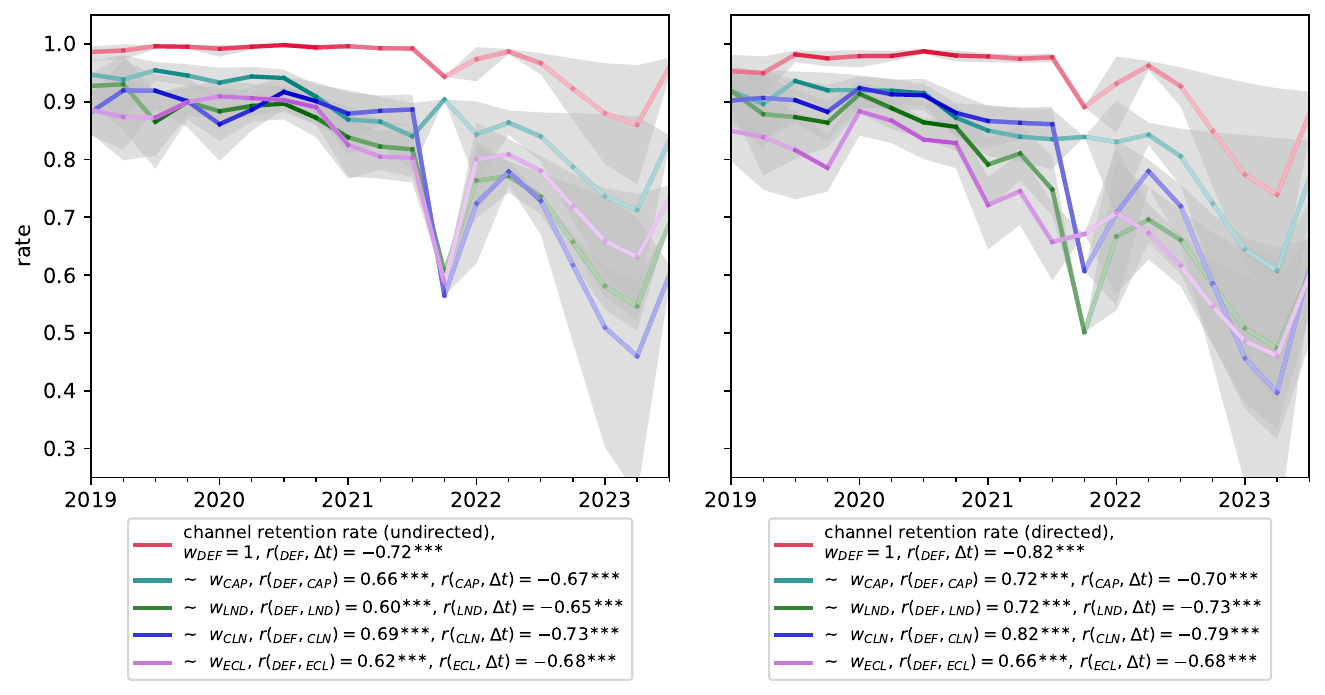}
\caption{The channel retention rate over time.}\label{fig13}
\begin{tablenotes}[]\footnotesize\item[]
Note. Quarterly averaged estimates (solid lines) are presented together with bootstrapped ($n = 100$) 95\% confidence intervals (shaded areas). Line colour intensity reflects the normalized data coverage for each time period based on the available snapshots. The channel retention rate $(I_{channel})$ and its weighted complement $(I_{channel}^{w})$ are calculated as defined in Eqs. \eqref{eq7} and \eqref{eq10} using a bootstrap technique (see \textit{Appendix A}).
\end{tablenotes}
\end{figure*}

To further examine operational stability, reflecting the observed cost-based feasibility of channels within the evolving network, we compute \textit{channel retention rates} under different payment pathfinding strategies (see \textit{Figure~\ref{fig13}}). Because the computational effort required for this analysis is highly intensive (consider all the combinations of retained channels and nodes), we rely on a bootstrapped version of the metric. This approximation is justified by calculating the \textit{RMSE} between the bootstrapped estimates and their exact computations, comprising measurements using both directed and undirected graph specifications with different cost functions (see \textit{Appendix A}). The resulting total \textit{RMSE} is reasonably low (0.007), indicating that the trade-off between the exact and bootstrapped metrics is justified for this analysis, which focuses primarily on temporal trends and confidence intervals.

\begin{figure*}[t]
\centering
\hspace{-2.5cm}
\includegraphics[scale=0.5, center]{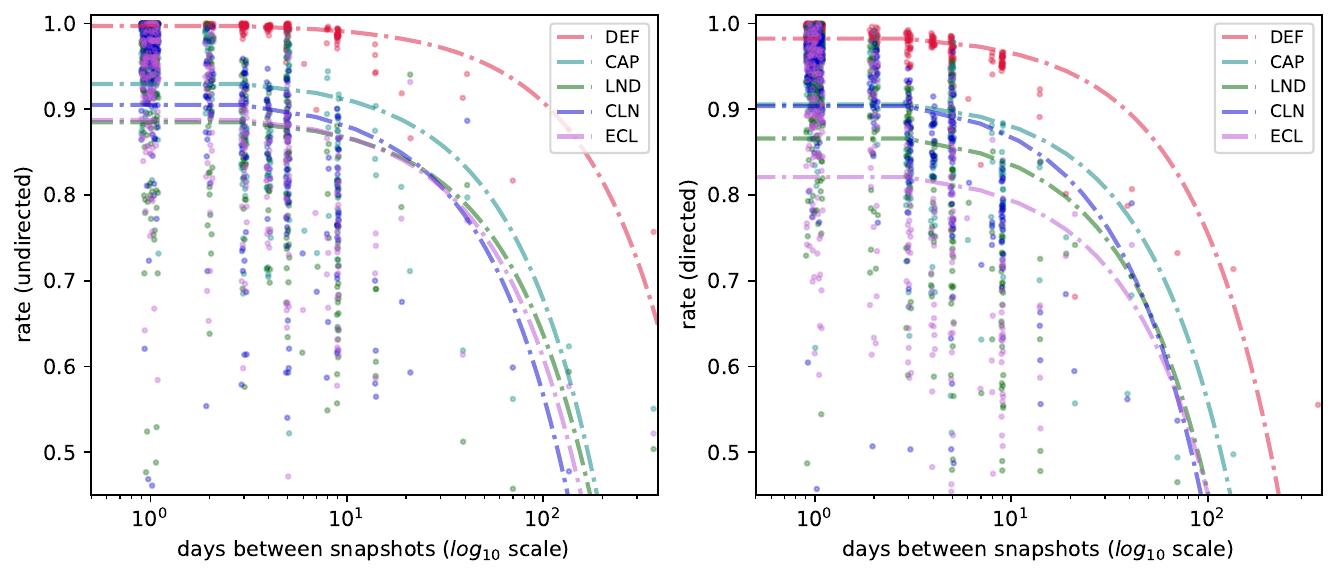}
\caption{Relationship between channel retention rates and elapsed time.}\label{fig15}
\begin{tablenotes}[]\footnotesize\item[]
Note. Lines represent robust linear regression fits relating channel retention rates to the time interval in days between snapshots.
\end{tablenotes}
\end{figure*}

The resulting average retention rates are consistently high, with most values exceeding 0.7. Furthermore, the different weighting methods exhibit similar temporal trends. The Spearman's correlation between retention rates and the time interval in days between snapshots is, on average, -0.7, indicating that retention rates effectively capture network evolution over time (see \textit{Figure~\ref{fig13}}). This suggests that the observed rates are influenced not only by the timing of the measurements but also by underlying network dynamics, whose intensity may vary across different periods. To further illustrate the relationship between channel retention rates and time across the different pathfinding strategies, we fitted robust linear regression models (see \textit{Figure~\ref{fig15}}). These results also confirmed the negative relationships highlighted by the correlation analyses above.

These findings also indicate that, although the previous analysis highlights a gradual isolation of node communities, observable transactional stability remains significantly unaffected. The exact overlap between the earliest available snapshot (20-01-2019) and the most recent one (16-07-2023) further underscores this persistence: approx. at least 73\% of equivalent channels are retained across the considered payment pathfinding strategies. This observation is supported by available independent statistics indicating that the average channel age as of 25 February 2026 is 730.1 days (approx. two years, see https://1ml.com/statistics).

Thus, high weighted retention rates indicate that LN routing operates on a merely stable backbone, even when superficial topological properties (e.g., clustering) evolve. \textit{Figures~\ref{fig12}} and \textit{\ref{fig13}} together demonstrate observed operational robustness. Even under realistic routing cost functions, equivalent payment paths exhibit strong retention, suggesting that routing behaviour remains stable despite gradual topological reconfiguration.

As a side note, while this reflects the current observed operational robustness of the network, such stability is not guaranteed in the future. The ongoing loss of information centrality and the growing number of communities may eventually reach a critical threshold at which payment flows might be more substantially impacted.

Finally, to comprehensively quantify the decentralization and inequality within the LN, we employ the \textit{Gini index}, following the approach of Zabka et al.\citep{Zabka2024}. The Gini index provides a visual representation of betweenness centrality inequality, illustrating the extent to which a small subset of nodes dominates the routing of payments therefore reflecting overall transactional stability.

\begin{figure*}[t]
\centering
\hspace{-2.5cm}
\includegraphics[scale=0.5, center]{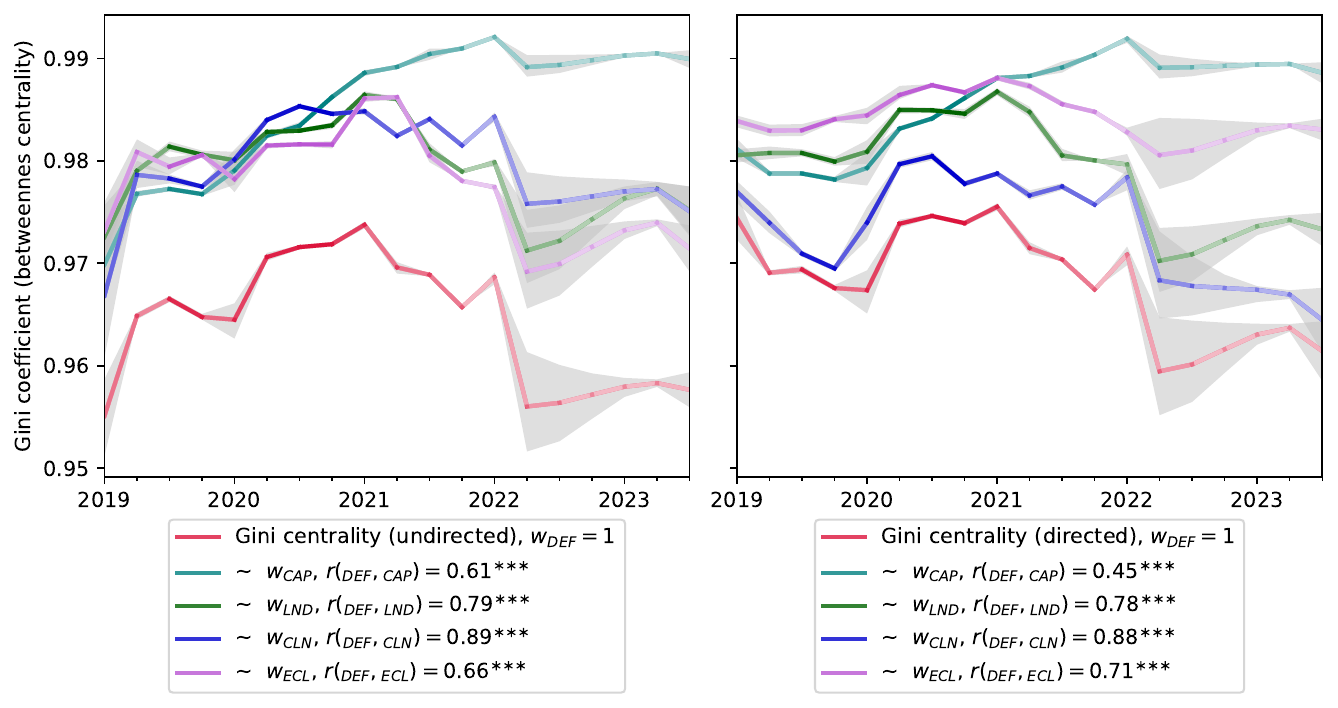}
\caption{Gini coefficients for the network betweenness centrality over time.}\label{fig14}
\begin{tablenotes}[]\footnotesize\item[]
Note. Quarterly averaged estimates (solid lines) are presented together with bootstrapped ($n = 100$) 95\% confidence intervals (shaded areas). Line colour intensity reflects the normalized data coverage for each time period based on the available snapshots. All calculations are performed for both undirected (left panel) and directed (right panel) graph representations, considering the unweighted case ($w_{DEF} = 1$), the capacity-weighted specification ($w_{CAP}$), and realistic pathfinding strategies ($w_{LND}$, $w_{CLN}$, $w_{ECL}$).
\end{tablenotes}
\end{figure*}

A lower Gini index indicates a more equitable and decentralized network structure, whereas a higher value reflects increasing centralization and heightened vulnerability -- particularly if the most frequently used nodes become unresponsive. A topology dominated by a small number of highly influential nodes, through which the majority of payments are routed, may undermine the overall robustness of the network\citep{Zabka2024}. However, from the perspective of transactional stability, if these centralized nodes remain online for extended periods or can be seamlessly replaced, together with their associated payment channels -- pathfinding stability can be preserved. 

As illustrated in \textit{Figure~\ref{fig14}}, the LN stay centralized over time, with Gini values exceeding 0.95 and subsequently oscillating above this threshold. Gini values above 0.95 indicate extreme concentration, comparable to highly centralized infrastructure networks. This pattern persists even when accounting for realistic pathfinding strategies, whose dynamics appear to be strongly positively correlated with unweighted Gini measure. 

On the one hand, our findings confirm the results reported in Zabka et al.\citep{Zabka2024}, indicating a high degree of centralization. On the other hand, our results extend this empirical evidence by demonstrating that the network remains operationally stable, despite the presence of central nodes that may be regarded as structural vulnerabilities in generic case.

However, this concentration should be interpreted as a potential vulnerability rather than direct evidence of systemic fragility. Demonstrating robustness under stress would require dedicated attack simulations, targeted hub-removal experiments, correlated failure scenarios, or liquidity shock analyses, which are outside the scope of the present study.

Overall, this analysis supports the conclusion that in general the LN remains largely operationally stable, even as it gradually evolves over time. Despite increasing centralisation, there is evidence that LN routing algorithms explicitly bias toward reliable, high-availability nodes, mitigating some of the operational risks implied by high centrality inequality. However, this reliance may increase correlated systemic risk if major hubs fail.

The results therefore reveal a structurally reconfiguring yet operationally persistent payment system: while mesoscopic cohesion (clustering, community overlap, global efficiency etc.) gradually declines, the distributional backbone and routing-equivalent channel structure remain remarkably stable. Thus, LN fragmentation reflects structural sparsification rather than systemic instability.

\section{Discussion}

This study provides a longitudinal, multi-perspective analysis of the Lightning Network using reconstructed topology snapshots spanning 2019–2023. The results reveal a network that is simultaneously evolving structurally and persisting operationally. While several mesoscopic cohesion metrics (clustering, transitivity, global efficiency etc.) exhibit gradual decline, distributional characteristics and routing-equivalent channel structures remain stable over time.

\subsection{Structural Evolution and Fragmentation}

Our results confirm and extend prior findings by \cite{Zabka2024} and \cite{Rohrer2019}. The LN continues to exhibit heavy-tailed degree distributions consistent with hub dominance, persistent sparsity, and high betweenness centralization. However, the present analysis adds a temporal dimension showing that sparsification is not merely static but progressive.

Declining density, clustering, and transitivity suggest reduced local redundancy and increasing structural fragmentation. At the same time, the growing number of detected communities indicates increasing modularization. Importantly, this fragmentation does not manifest as random disintegration but as structured sparsification: the backbone of high-degree, high-capacity nodes remains intact while peripheral nodes connect selectively. This pattern is consistent with economic channel-opening incentives described theoretically by Guasoni et al.\citep{Guasoni2024}, where cost minimization mechanisms naturally produce hub-oriented topologies.

Note that although the observed increase in detected communities is consistent with increasing modularization, community counts are influenced by both network size and the resolution characteristics of modularity-based algorithms. Therefore, the present analysis cannot distinguish entirely between genuine fragmentation and scale-related effects.

\subsection{Multi-Scale Stability: Structural vs. Operational}

A central contribution of this paper is the distinction between three forms of stability: (i) structural (topological cohesion), (ii) technical (node and channel retention), and (iii) operational (observed weighted routing feasibility).

While structural cohesion decreases at the mesoscopic level, technical and distributional stability remains high. Kolmogorov–Smirnov statistics and Wasserstein distances indicate that degree and shared capacity distributions change only minimally between successive snapshots. This suggests that the technical backbone of the network is preserved even as local structure evolves.

More importantly, observable operational stability measures reveal that routing-equivalent channels are retained between successive snapshots, across all examined pathfinding strategies (LND, CLN, ECL). Thus, despite superficial topological change, the LN maintains functional continuity for payment routing. In this sense, LN evolution reflects structural reconfiguration rather than systemic instability.

It is important to note that the stability analysis presented in this study is based exclusively on the publicly observable LN topology derived from the gossip protocol, including publicly advertised channels and channel capacities. It does not capture private channels, actual channel balances, or hidden liquidity constraints, all of which influence real-world payment routing. Consequently, the operational stability measures reported here should be interpreted as indicators of observable topology-based routing feasibility, representing the best possible assessment that can be inferred from publicly available network information rather than direct evidence of end-to-end payment stability.

\subsection{Centralization and Systemic Risk}

The Gini analysis confirms extreme inequality in betweenness centrality, with values persistently exceeding 0.95. These results corroborate prior evidence of routing centralization in the LN\citep{Zabka2024}. However, centralization can enhance short-term routing efficiency by concentrating liquidity and reliability in a stable set of nodes. We have demonstrated that LN routing implementations such as LND, CLN, and ECL inherently bias toward reliable, high-availability channels, reinforcing this effect. 

However, such concentration may increase correlated systemic risk\citep{SUHrisk}. If major hubs fail, become adversarial, or are subject to regulatory intervention, a large fraction of payment flows may be simultaneously affected. Thus, operational stability observed in historical data does not eliminate potential future fragility under stress scenarios.

It should also be noted, that shortest-path–based routing tends to concentrate flow along geodesic routes, which may overestimate hub dominance relative to real-world LN behaviour. In practice, multi-path payments and probabilistic pathfinding may distribute flows across multiple (even not optimal) alternatives. Although we took realistic pathfinding algorithms into account, the stochastic nature of routing was not fully accounted for due to reproducibility requirements. Therefore, the Gini coefficients should be interpreted as proxies for centralisation rather than accurate estimates of exact behaviour.

\subsection{Sensitivity and Representation Robustness}

To assess the robustness of our findings, we systematically compared metric behaviour under directed, undirected, unweighted, capacity-weighted and cost-weighted specifications. While absolute magnitudes vary across representations, particularly for betweenness and information centrality, clustering and communities, the temporal trends and relative stability patterns remain highly consistent. Capacity-weighted and routing-aware representations preserve the persistence of the backbone structure and confirm that retained channels account for a substantial share of network capacity, indicating that stability is not merely topological but economically meaningful. Directed representations introduce asymmetry effects that moderately affect centrality rankings and path-length distributions; however, these adjustments do not alter the principal conclusions regarding sparsification, concentration, and high retention rates. The undirected and unweighted abstractions, therefore, remain analytically sound for capturing macro-structural evolution and distributional stability, as they approximate the dominant connectivity backbone of the network. Nonetheless, direction- and weight-aware specifications are essential for interpreting routing dominance and economic influence. Taken together, the consistency of results across representation choices suggests that the reported structural and temporal stability properties are not artefacts of a particular graph representation but reflect persistent characteristics of the LN.

\subsection{Implications for Developers and Researchers}

The findings have several practical implications:
\begin{itemize}
    \item High backbone persistence justifies heuristic and reinforcement-learning–based routing approaches that rely on historical structural consistency.
    \item Persistent hub dominance implies that capacity allocation remains concentrated, potentially affecting fee competition and economic decentralization.
    \item Declining mesoscopic cohesion suggests that long-term fragmentation should be monitored to prevent thresholds where functional stability might degrade.
\end{itemize}

More broadly, this analysis and publicly available dataset provide reproducible reference metrics for benchmarking LN research, addressing a long-standing empirical gap in payment channel network analysis.

\section{Limitations}

Despite the breadth and depth of the analysed data, several limitations must be acknowledged. 

First, the analysis relies on publicly available gossip messages. Private changes and non-broadcasted modifications are not observable. Therefore, reconstructed topologies approximate the publicly visible LN rather than the complete operational network. While cross-validation against external statistics suggests high fidelity, hidden balances and private routing strategies remain unaccounted for. 

Second, snapshot density varies across the observation time frame due to historical archival gaps. Although bootstrapped confidence intervals and explicit coverage visualization partially help with this issue, periods of sparse data may exhibit higher uncertainty.

Third, weighted analyses rely on documented cost functions of major LN implementations. However, real-world routing behaviour includes dynamic heuristics and fuzziness, liquidity constraints, channel balance uncertainty, and adaptive retry strategies. Consequently, operational stability metrics approximate routing feasibility rather than replicate exact payment behaviour. However, this is a reasonable limitation necessary to ensure the reproducibility of results.

Fourth, the increase in detected communities may partially reflect algorithmic resolution limits inherent in modularity-based methods. As network size increases, community counts may rise mechanically. Therefore, fragmentation results should be interpreted with this methodological caveat in mind.

Fifth, the present study evaluates historical evolution but does not simulate adversarial attacks, correlated and systemic failures, or liquidity shocks. Thus, conclusions regarding systemic robustness apply to observed dynamics rather than worst-case scenarios.

\section{Conclusion}

This paper presents the most comprehensive longitudinal topology analysis of the Lightning Network to date, based on 336 reconstructed snapshots spanning five years. By computing an extensive set of network-science metrics under directed, undirected, weighted, and routing-aware representations, we provide a multi-scale characterization of LN structural and temporal evolution.

The results reveal a network undergoing gradual structural sparsification and increasing modularization while preserving strong distributional and observed operational stability. Degree and capacity distributions remain highly stable across observed time frame, node and channel retention rates are persistently high, and routing-equivalent paths are maintained under realistic pathfinding cost models. This supports the claim that the network exhibits persistent observable routing characteristics under historically observed topology evolution.

At the same time, extreme centralization in routing betweenness persists, raising questions regarding long-term decentralization and correlated systemic risk.

Overall, the LN appears structurally reconfiguring yet operationally persistent: fragmentation at the mesoscopic level does not currently translate into functional instability. These findings contribute both empirically and methodologically to payment channel network research and provide a validated benchmark dataset for future studies on routing optimization, robustness analysis, and decentralized infrastructure design.

Future research should extend this framework toward stress-testing scenarios, liquidity-aware simulation, and incentive-compatible topology design to assess how stable the Lightning Network remains under adversarial or high-load conditions.

\section*{Declaration of competing interest}
The authors declare that they have no known competing financial interests or personal relationships that could have appeared to influence the work reported in this paper.

\section*{Data availability}
The raw data is publicly available and referenced in the paper. The resulting data is documented, provided in a \textit{.csv} file, and available in the public repository at \url{https://github.com/ellariel/ln-comprehensive-analysis} to facilitate further analysis.

\section*{Authorship contribution statement}
DV: Writing – review \& editing, Writing – original draft, Visualization, Software, Methodology, Data curation, Conceptualization. JMG: Writing – review \& editing, Validation, Supervision.

\section*{Funding}
No funding received for this research.

\section*{Appendix A -- Testing the bootstrapped estimates}

Table below reports the \textit{RMSE} and Spearman's correlation ($r$) between the exact retention rates and their bootstrapped estimates across different graph representations and weighting strategies. The bootstrapped estimates were computed as the median over 100 resamples, each consisting of 100 randomly selected overlapping nodes (as defined in Eqs. \eqref{eq7} and \eqref{eq10}) with replacement for each graph snapshot.

\begin{table}[!htbp]
\centering
\label{tab:bootstrap_validation}
\begin{tabular}{lccccc}
\hline
Graph / Weighting & DEF & CAP & LND \\
\hline
Directed &
\begin{tabular}[c]{@{}c@{}}
$RMSE = 0.012, n = 222$\\
$r = 0.935, p < 0.001$
\end{tabular} &
\begin{tabular}[c]{@{}c@{}}
$RMSE = 0.007, n = 131$\\
$r = 0.991, p < 0.001$
\end{tabular} &
-- \\
\hline
Undirected &
\begin{tabular}[c]{@{}c@{}}
$RMSE = 0.002, n = 331$\\
$r = 0.970, p < 0.001$
\end{tabular} &
\begin{tabular}[c]{@{}c@{}}
$RMSE = 0.003, n = 175$\\
$\rho = 0.998, p < 0.001$
\end{tabular} &
\begin{tabular}[c]{@{}c@{}}
$RMSE = 0.005, n = 314$\\
$r = 0.998, p < 0.001$
\end{tabular} \\
\hline
\end{tabular}
\end{table}

The results show consistently low total \textit{RMSE} values ($RMSE_{total} = 0.007, n = 1173$) and high Spearman's correlations ($r_{total} = 0.980, p < 0.0001$), indicating strong agreement between the exact and bootstrapped computations. These findings suggest that the bootstrapped estimates serve as reliable proxies for the exact metrics, providing a well-justified trade-off between computational efficiency and accuracy.


\bibliography{sn-bibliography}

\end{document}